\pdfoutput=1

\documentclass[11pt]{article}

\usepackage[final]{acl}

\usepackage{times}
\usepackage{latexsym}
\usepackage[T1]{fontenc}
\usepackage[utf8]{inputenc}
\usepackage{microtype}
\usepackage{inconsolata}
\usepackage{graphicx}
\usepackage{multirow}
\usepackage{booktabs}
\usepackage{calc}
\usepackage{subfig}
\usepackage{xparse}
\usepackage{enumitem}
\usepackage{listings,tikz}
\usepackage[most]{tcolorbox}
\usepackage[frozencache,cachedir=minted-cache]{minted}
\tcbuselibrary{minted}
\usepackage{makecell}

\newtcblisting{pythoncode}{
  listing engine=minted,
  breakable,
  listing only,
  nobeforeafter,
  minted style=colorful,
  minted language=python,
  enhanced,
  minted options = {
        breaklines=true, 
        fontsize=\small
      },
}

\newtcblisting{javacode}{
  listing engine=minted,
  breakable,
  listing only,
  nobeforeafter,
  minted style=colorful,
  minted language=java,
  enhanced,
  minted options = {
        breaklines=true, 
        fontsize=\small
      },
}

\usepackage[T1]{fontenc}

\usepackage[utf8]{inputenc}

\usepackage{microtype}

\usepackage{inconsolata}

\definecolor{dkgreen}{rgb}{0,0.6,0}
\definecolor{gray}{rgb}{0.5,0.5,0.5}
\definecolor{mauve}{rgb}{0.58,0,0.82}

\title{CodeComplex: Dataset for Worst-Case Time Complexity Prediction}

\author{Seung-Yeop Baik\\
  Yonsei University \\
  Seoul, South Korea \\\And
Joonghyuk Hahn\\
   Yonsei University \\
  Seoul, South Korea \And
  Jungin Kim \\
   Yonsei University \\
  Seoul, South Korea \And
  Aditi \\
University of Seoul \\
 Seoul, South Korea \AND
  Mingi Jeon \\
  Kangwon National University \\
  Chuncheon, South Korea \And
  Yo-Sub Han\\
  Yonsei University \\
  Seoul, South Korea \And
  Sang-Ki Ko\\
  University of Seoul \\
 Seoul, South Korea
  }

\begin{document}
\maketitle

\begin{abstract}
Reasoning ability of Large Language Models (LLMs) is a crucial ability, 
especially in complex decision-making tasks. 
One significant task to show LLMs' 
reasoning capability is code time complexity prediction, 
which involves various intricate factors 
such as the input range of variables and conditional loops. 
Current benchmarks fall short of providing 
a rigorous assessment due to limited data, 
language constraints, and insufficient labeling. 
They do not consider time complexity based on input representation 
and merely evaluate whether predictions fall into the same class, 
lacking a measure of how close incorrect predictions are to the correct ones.
To address these dependencies, we introduce CodeComplex, 
the first robust and extensive dataset designed 
to evaluate LLMs' reasoning abilities in predicting code time complexity.
CodeComplex comprises 4,900 Java codes and an equivalent number of Python codes,
overcoming language and labeling constraints,
carefully annotated with complexity labels based on input 
characteristics by a panel of algorithmic experts.
Additionally, we propose specialized evaluation metrics
for the reasoning of complexity prediction tasks, 
offering a more precise and reliable assessment 
of LLMs' reasoning capabilities.
We release our dataset and baseline models\footnote{https://github.com/sybaik1/CodeComplex} publicly to encourage the relevant (NLP, SE, and PL) communities to utilize and participate in this research.

\end{abstract}

\section{Introduction}


Large Language Models (LLMs) demonstrate significant potential 
in complex decision-making tasks, 
with their inference capabilities being particularly valuable 
in software development~\cite{Austin21,JainVINPRS21}. 
To further test LLM's inferencing abilities,
we showcase this domain of predicting the code time complexity, 
which requires the consideration of numerous intricate factors. 
For example, factors such as algorithmic structure~\cite{Turing36,BentleyHS80}, 
data input size, and resource constraints can all influence the time complexity of code~\cite{Nogueira12,HutterXHL14}. 
Understanding and optimizing these factors is crucial 
for improving the performance of complex algorithms and generating efficient code~\cite{PengZLKHL21,LuGRHSBCDJTLZSZ21}.

Despite the existence of benchmarks for the time complexity analysis,
such as CoRCoD~\cite{SikkaSKUSZ20}, 
and TASTY~\cite{moudgalya2023tasty}, these benchmarks have limitations on their current state.
Notably, CoRCoD is the only publicly available dataset that is small in size, and while TASTY considers both time and space complexity, its dataset remains undisclosed. 
Consequently, there is a clear need for a comprehensive and publicly accessible benchmark that addresses these shortcomings.

Our work aims to fill this gap by introducing CodeComplex, 
a dataset designed to be the definitive benchmark 
for evaluating LLMs' time complexity inference abilities. 
CodeComplex offers several distinct advantages over existing benchmarks. 
Firstly, it provides a larger and more diverse dataset, 
encompassing a broad range of programming languages 
beyond the limited scope of CoRCoD, which focuses solely on Java. 
Secondly, our dataset encompasses a more comprehensive set of labeled complexity classes that cover general-use problem-solving algorithms, enabling a more detailed analysis.
Thirdly, CodeComplex considers the representation of input data, distinguishing between numeric values and input size indicators, which is critical for accurate complexity analysis. 
Lastly, we suggest detailed metrics that allow for a more nuanced assessment of LLMs' performance, moving beyond simple class-based evaluations.

In summary, our contributions are as follows:
\begin{enumerate}
    \item Comprehensive Dataset: We present a novel dataset with a comprehensive range of algorithmic problems that includes bilingual programming languages,
    offering a significant expansion upon existing benchmarks.
    \item Detailed Complexity Analysis: We annotate the code complexity through complexity analysis of the representation of input data,
    distinguishing between numeric interpretations and input size indicators.
    \item Robustness of Evaluation Metrics: We implement precise evaluation metrics for a rigorous and nuanced assessment of LLMs' time complexity inference, facilitating accurate comparisons with state-of-the-art baselines.
\end{enumerate}

Through these contributions, CodeComplex aims to advance the study
of LLMs inference capabilities with the task of predicting code time 
complexity, 
ultimately fostering the development of optimized software.
Our benchmark sets a new standard for the evaluation of LLMs, 
providing a robust and comprehensive tool for researchers and practitioners alike.


\begin{figure*}[!t]
    \centerline{\includegraphics[width=\textwidth]{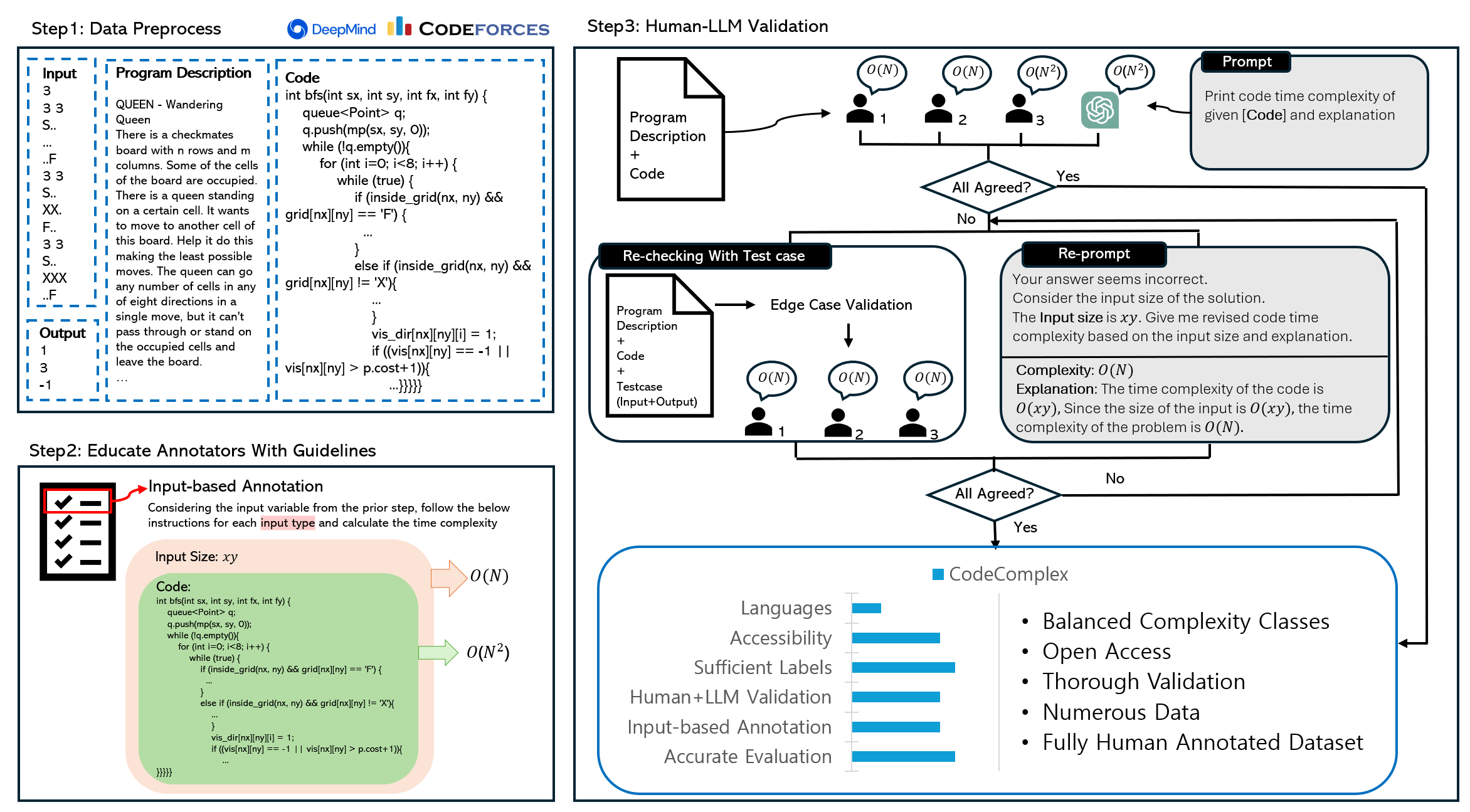}}
    \caption{Overview of the CodeComplex dataset creation process.}
    \label{fig:overview}
\end{figure*}

\section{Related Work}
The LLMs have led to numerous advances in the field of natural language processing~\cite{BrownMRSKDNSSAA20,ChowdheryNDBMRBCSGSSTMRBTSPRDHPBAI23}.
Therefore, recent research is ongoing to improve the reasoning ability of LLMs.
Numerous prompt engineering techniques emerged, such as zero-shot, few-shot, chain of thought, and prompt changing~\cite{Wei0SBIXCLZ22,WangWSLCNCZ23,YaoYZSGCN23}.
Specifically, the chain of thought prompting was proposed to encourage them to engage in inferential thinking, with the subsequent application of arithmetic,
commonsense, and sentiment reasoning.
The methodologies demonstrated that when the LLMs' reasoning was improved, their capability to perform a range of tasks was enhanced.

Nevertheless, it is evident that LLMs are still constrained in their capacity to reason about complex tasks.
The enhancement of LLM's capacity to reason effectively in complex domains is highly dependent upon the availability of a domain-specific and sophisticated dataset.
In the field of code time complexity, some representative complex tasks have been proposed as a method for improving the reasoning ability of LLM models.
Recently, \citet{SikkaSKUSZ20} explored code complexity prediction using machine learning-based methods. 
They curated the CoRCoD dataset comprising 929 annotated Java codes. 
These codes were enriched with various hand-engineered features extracted from the code, 
encompassing counts of loops, methods, variables, jumps, breaks, switches, and the identification of specific data structures or algorithms like priority queues, hash maps, hash sets, and sorting functions. 
Employing machine learning classification algorithms such as $K$-means, random forest, decision tree, SVM, and more, they made predictions based on these diverse features. 
Additionally, they explored graph2vec~\cite{NarayananCVCLJ17}, a neural graph embedding framework that operates on a program's AST and achieves comparable performance results.

Another exploration by \citet{PrennerR21} scrutinized the potential of pre-trained programming language understanding models, 
particularly CodeBERT~\cite{FengGTDFGS0LJZ20}, for predicting code complexity. Their experiments showcased promising results, suggesting that pre-trained models could serve as a viable solution in this domain.
In the most recent development, \citet{moudgalya2023tasty} tackled the analysis of time and space complexity using language models. They leveraged codes sourced from GeeksForGeeks\footnote{\url{https://www.geeksforgeeks.org/}} and CoRCoD, alongside a dataset comprising 3,803 Java codes. Their work showcased the viability of fine-tuning pre-trained language models such as GraphCodeBERT~\cite{GuoRLFT0ZDSFTDC21} for predicting both time and space complexity, thereby opening new avenues for exploration in this field.

\section{The CodeComplex Dataset}

The CodeComplex dataset contains a collection of codes
written in two languages, Java and Python,
from a competitive programming platform.
Our dataset originates from Codeforces and collects data from CodeContests~\cite{Li22AlphaCode}, a competitive programming dataset tailored for machine learning applications created by DeepMind. It comprises 9,800 codes, evenly split between Java and Python, with 4,900 codes each. We have categorized these codes into seven distinct complexity classes: constant ($O(1)$), linear ($O(n)$), quadratic ($O(n^2)$), cubic ($O(n^3)$), logarithmic ($O(\ln n)$, $O(n \ln n)$), and exponential. Each class contains a minimum of 500 Java and Python codes.

We annotated all 9,800 codes with experts, which include the 
317 Java codes in the CoRCoD dataset from Codeforces.
It is worth mentioning that the CoRCoD, 
a previous dataset used for code complexity prediction, 
categorizes Java codes into five complexity classes: 
$O(1)$, $O(n)$, $O(n^2)$, $O(\ln n)$, and $O(n \ln n)$. 
However, it suffers from imbalanced class distribution, 
evident in Table~\ref{tab: statistic}, 
with a relatively small size of 929 Java code samples in total. 
Our expansion significantly enhances the dataset's value for research, 
particularly concerning DL-based models outlined in Section~\ref{ssec: experiments}.

\begin{table}[!htbp]
\setlength{\tabcolsep}{0.4em}
\renewcommand{\arraystretch}{0.8}
\centering
\begin{tabular}{l|c|cc}
\toprule
\multirow{2}{*}{\bf Class} & {\bf CoRCoD} & \multicolumn{2}{c}{\bf CodeComplex}\\\cmidrule{2-4}
 & {\bf Java} & {\bf Java} & {\bf Python} \\
\midrule
${O(1)}$ & 143 & 750 (+ 62) & 791\\
${O(n)}$ & 382 & 779 (+ 117) & 853\\
${O(n^2)}$  &  200 & 765 (+ 48) & 657\\
${O(n^3)}$  & 0 & 601 & 606\\
${O(\ln n)}$  & 54 & 700 (+ 18) & 669\\
${O(n\ln n)}$  & 150 & 700 (+ 72) & 796\\
exponential  & 0  & 605 & 528\\
\midrule
{\bf Total}  & 929  & 4,900 (+ 317) & 4,900\\
\bottomrule
\end{tabular}
\caption{Statistical difference between CoRCoD and CodeComplex. Numbers in parentheses imply the number of codes from CoRCoD.}
\label{tab: statistic}
\end{table}

\subsection{Data Collection}\label{se:collection}


The original corpus of code is from CodeContests, which collected 128 million codes from Codeforces.
The corpus only contained information about the contest ID, problem, username, language, acceptance, and statistics (runtime and memory).
We extracted the selected problems from this corpus and identified 
each code's complexity.

Code samples were selected within the matching candidates with the following conditions.
First, we checked the relevance of the problem.
There are many problems within a coding competition, but not all of them fall into the scope of 
complexities we seek to compromise. Therefore, the problems were first analyzed to check 
whether or not they were in the complexity class of our dataset.
If the problem was determined to be in one of the seven complexity classes, then we marked the problem 
as a candidate for the dataset.
This helps to establish a clear base dataset for the complexity domain.
Second, we checked the completeness and correctness of the code.
We filtered codes that are available to pass the given problem in the contest,
meaning that the code is functional, self-contained, and correct on the given task.
One of the reasons for using code competition data is that 
we can check if the code is correct for the problem.
Lastly, we wanted a large pool of code samples for a given problem.
We took code samples from problems with abundant submissions.
This helped to clarify the problem's robustness and variation.

Consider the following Python program that solves a problem with $O(1)$ time complexity:
\begin{pythoncode}
buf = input()
hand = buf.split()
t = []
for i in range(3):
    t.append([])
    for j in range(9):  
        t[i].append(0)
for x in hand:
    idx = 0
    # Following lines are omitted.
\end{pythoncode}

Despite the short length of the code, it is not trivial to understand that the time complexity of the above code is actually constant, which implies that the number of instructions for executing the program does not depend on the input size. In fact, the problem description says that the input always consists of three strings separated by whitespace and, therefore, the size of the list \textsf{hand} is actually constant. Hence, it is impossible to correctly calculate the time complexity of a code only by analyzing the code, as the problem description sometimes has a big hint to determine the time complexity.


\subsection{Data Preprocessing}
Data preprocessing is an important step in preparing datasets for analysis or machine learning tasks.
In this process, we utilize {\em dead code elimination} and {\em comment removal}.
Dead code elimination involves removing any code that does not contribute 
to the functionality or output of the program, thereby reducing unnecessary 
clutter.
From each code, we marked irrelevant codes and unreachable codes as dead codes.
Irrelevant code involves variables, functions, and classes that were never used or never called,
and unreachable code involves conditional statements that cannot be satisfied
and statements that cannot be reached because of control statements such as continue and return.

On the other hand, comment removal entails stripping out any comments within the codebase,
which are meant for human understanding. 
We removed the comments since the fragments could be exploited by the models 
to improve the accuracy of predicting the time complexity of models.

\subsection{Annotation Process}

Our primary objective is to create a solid foundation for accurately classifying time complexities. 
To achieve this, we have meticulously designed a procedure to generate a robust dataset with minimal noise and high quality.
We specifically filter `correct' Java and Python codes, ensuring they pass all test cases, including hidden ones. 
These codes form the basis of our statistical population.
Categorizing problems based on problem-solving strategies involves leveraging annotations from CodeContests.
Each problem in the dataset is associated with a plausible problem-solving strategy, such as brute force, dynamic programming, or backtracking, as outlined in CodeContests.
Following this initial categorization, a detailed analysis of each problem is conducted. This analysis considers input and output variables, utilized data structures, and the overall workflow of the code. Subsequently, the code for each problem is annotated based on its specific input characteristics. More precisely, we take the largest input variable as the main factor in calculating the overall time complexity. By analyzing the code, we consider each control sequence on the code to determine if the input impacts a control segment or is constant. Note that we assume a {\em unit-cost RAM model} that requires the same cost for accessing all memory locations for calculating the time complexity.
Our core annotation process adheres to four key rules:
\begin{enumerate}
    \item Consider the input size and the output size as parameters to determine time complexity, with measurement based on the largest parameter among the input variables.
    \item Account the impact of used packages and libraries, such as hashmap, sorting, and string-matching algorithms, on time complexity.
    \item Treating each test case within a single input separately for complexity measurement.
    \item Classifying cases with fixed constants as having a constant time complexity.
\end{enumerate}
The annotation was held by three annotators who have expertise in the algorithm.
In the initial annotation step, each annotator annotated each problem 
independently. 
Each reasoned on how we judged the input and annotated the time complexity.

\begin{figure*}[!thb]
    \centering
    \includegraphics[width=.96\textwidth]{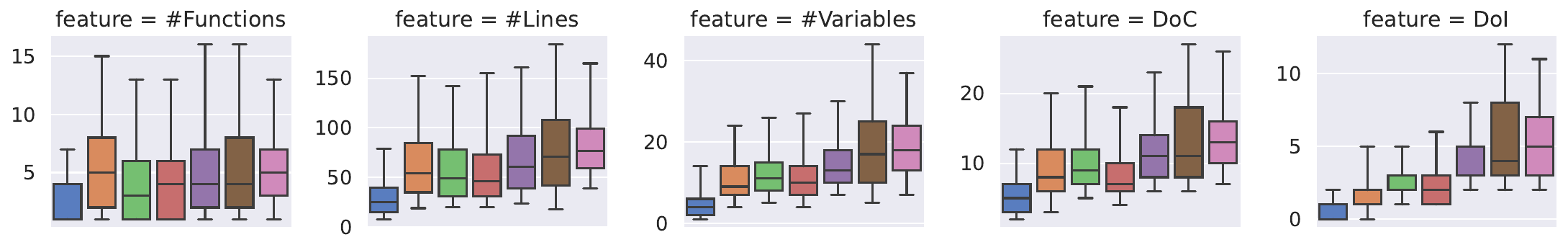}
    \includegraphics[width=.96\textwidth]{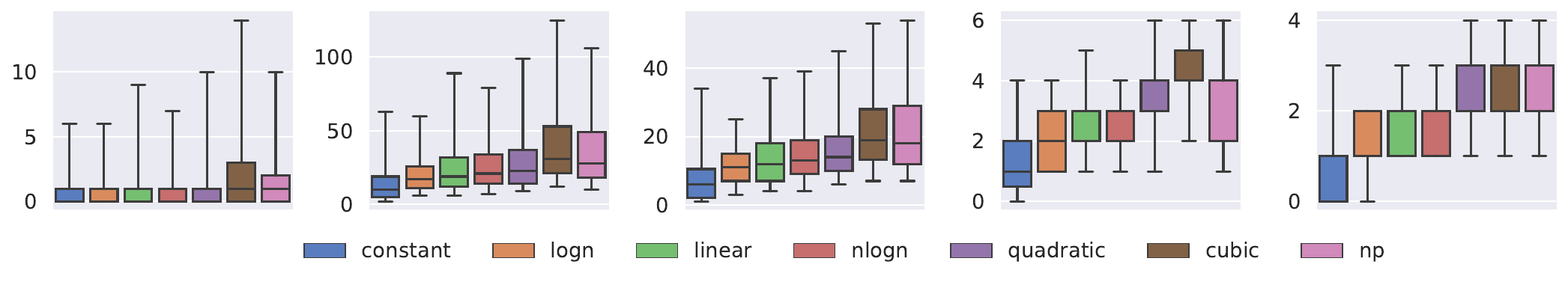}
    \caption{Statistics of CodeComplex dataset. The first and second lines are for Java and Python codes, respectively.}
    \label{fig:features}
\end{figure*}

During the agreement process, the annotators collaborated closely to reconcile any discrepancies in their annotations. 
We engaged in thorough discussions, sharing our reasoning and insights to reach a consensus on the 
appropriate time complexity classification for each problem. 
In cases where disagreements arose, the annotators carefully evaluated the evidence 
and considered alternative perspectives before arriving at a mutually acceptable classification.
The involvement of ChatGPT serves as a neutral advisor to validate the annotations
and offer additional perspectives on complex cases.
Through open communication and collaborative decision-making, 
the annotators ensure the accuracy and reliability of the final dataset.
However, it is essential to note the significant impact of input formats 
and constraints on the actual time complexity of algorithmic problems. 
These constraints often lead to deviations from the ideal time complexity. 
Think of a scenario in which the input exploits the problem constraints in time complexity.
Despite the problem of having a quadratic time complexity, 
the provided input constraints may result in linear running time. 
Moreover, determining the parameter for complexity measurement becomes crucial
when faced with multiple input parameters. Additionally, certain code submissions
optimize execution based on problem constraints, thus influencing code complexity assessment.

\begin{table}[!htb]
\centering
\resizebox{\columnwidth}{!}{%
\begin{tabular}{lcccc}\hline
                       & TASTY & CODAIT & CoRCoD & CodeComplex \\ \hline
Languages              & O     & X      & X      & O           \\
Accessibility          & X     & X      & O      & O           \\
Sufficient Labels      & O     & $\bigtriangleup$     & $\bigtriangleup$     & O           \\
Input-based Annotation & X     & X      & X      & O          \\\hline
\end{tabular}%
}
\caption{Comparison between CodeComplex and other time complexity prediction datasets.}
\label{tab:compare_data}
\end{table}
The CodeComplex dataset offers a meticulously curated collection of algorithmic problems and corresponding Java and Python code submissions. It serves as a foundation for accurately classifying time complexities and problem-solving strategies.
Figure~\ref{fig:features} demonstrates basic statistics of the codes for the number of lines, functions, variables, depth of code~(DoC), and depth of iterations~(DoI).
Moreover, both Java and Python solutions displayed comparable characteristics when considering the depth of iterations, reflecting nested loops. However, a distinctive trait of Python code was its abundance of variables, potentially attributed to Python's lack of explicit variable declaration requirements. This inherent difference in variable declaration mechanisms might contribute to the observed discrepancy in variable counts between the two languages within our dataset. Table~\ref{tab:compare_data} summarizes the strengths of our dataset.
\subsection{Dataset Overview}
\paragraph{Various Data} We have collected the dataset to include multiple problems upon various algorithm categories.
Each complexity class includes problems from dynamic programming, brute force, dynamic programming, divide and conquer,
and much more. The dataset provides the problem tags from Codeforces, which the user can use to verify the algorithm categories.
Also, we can see in Figure~\ref{fig:features}, that our dataset is not biased in the number of functions or variables
for a given complexity class. The traditional models in Table~\ref{tab:JavaPython} fail to verify the complexity class from these features, which further
shows that the solution codes have an even distribution among the complexity classes.

\paragraph{Balanced Complexity Classes} One of the significant strengths of our dataset 
is the careful balancing of complexity classes. 
Balanced classes prevent bias in the reasoning process of LLMs 
and enable more accurate and generalized model performance 
by providing a comprehensive learning experience across all possible scenarios. 
However, in Tables~\ref{tab: statistic}, the CoRCoD dataset exhibits imbalances among the classes. 
In contrast, our dataset provides balanced class data, 
ensuring an accurate and reliable prediction result of LLM models.

\paragraph{Open Access} In previous research on benchmarking the time complexity of code, it was observed that open data sources for code complexity prediction are notably scarce, with CoRCoD being one of the few available datasets. In contrast, our dataset is an open-source resource, providing an accessible and practical solution for researchers and practitioners in the field.

\paragraph{Fully Human Generated and Annotated Dataset} The Codecomplex dataset is labeled manually by three human annotators 
and all final decisions and verifications are made by humans. 
Also, the source codes are collected from the codes before LLMs became present, 
which makes them free of machine-generated codes.

\paragraph{Thorough Validation} During the process of code complexity annotation, 
a thorough validation was conducted. 
Three annotators annotated using structured guidelines and cross-validated 
each other's annotations and the reasoning process.
Also, we used ChatGPT as a failsafe precaution to validate the labels.
If the response from ChatGPT disagrees, 
the annotators re-evaluated the code through edge case validation.

\section{Experiments} 
\label{ssec: experiments}
As a preliminary study on code complexity prediction using a large-scale dataset, we conduct experiments with well-known machine learning-based solutions and large language models(LLMs). First, we try to replicate the result by \citet{SikkaSKUSZ20} by employing traditional models such as decision tree (DT), random forest (RF), and support vector machine (SVM). Second, we use pre-trained programming language models (PLMs) such as CodeBERT~\cite{FengGTDFGS0LJZ20}, GraphCodeBERT~\cite{GuoRLFT0ZDSFTDC21}, UniXcoder~\cite{guo-etal-2022-unixcoder}, PLBART~\cite{ahmadCRC2021}, CodeT5~\cite{WangWJH2021}, and CodeT5+~\cite{WangWJH2021}. Note that we can further categorize these models into two groups where the first group (CodeBERT, GraphCodeBERT, and UniXcoder) only uses encoder architecture, and the second group (PLBART, CodeT5, and CodeT5+) exploits encoder-decoder architecture. Finally, we test the dataset on closed-source LLMs, ChatGPT3.5, ChatGPT4.0~\cite{OpenAI2024ChatGPT}, Gemini Pro~\cite{Google2024Gemini},
and test open source LLMs Llama~\cite{LLama2023}, CodeGemma, Gemma1 and Gemma2~\cite{gemma2024}, Mistral-Nemo~\cite{mistral7b2023}, Qwen2 and Qwen2.5~\cite{qwen22024}
from an instruction-tuned and a fine-tuned version of our dataset.

\subsection{Experimental Settings}
We divide the CodeComplex into training and test datasets by a 9 to 1 ratio for both Java and Python. 
As a result, the training and test datasets comprise 8,820 and 980 codes, respectively. Hyperparameters and methods to fine-tune each model can be found in section~\ref{hyperparameters}.

\subsection{Evaluation Metric}
\paragraph{Hierarchy Complexity Score(HC-Score)} 
There are multiple approaches for estimating complex reasoning 
capabilities of LLMs. To maximize the effectiveness of
our dataset, we propose a novel evaluation metric, 
the Hierarchy Complexity Score (HC-Score). 
Unlike traditional accuracy measures, 
the HC-Score incorporates considerations 
of the code-time-complexity hierarchy, 
penalizing predictions in proportion 
to their deviation from the correct answer. 
Our evaluation metric~{\rm HC-Score} is computed as follows:
\[
\mathrm{HC\text{-}Score}(P,R) = \frac{\sum_{i=1}^N \frac{|p_i - r_i|}{\text{Number of class}}}{N},
\]
where we have $N$ number of $p_i\in P$ and $r_i\in R$
where $p_i$ is a prediction and
$r_i$ is an answer. Also, we can expand this metric by reducing the scoring window of a single class as follows:
\[
\mathrm{HC\text{-}Score}_{w}(P,R) = \frac{\sum_{i=1}^N max(1-\frac{|p_i - r_i|}{w}, 0)}{N},
\]
where $w$ is the size of the window for each class.
Predictions that closely approximate the correct result incur minimal penalties, 
while those with substantial errors are penalized more heavily. 
This nuanced scoring system aims to provide a more precise assessment 
of an LLM's reasoning abilities by accounting for the complexity of the tasks at hand. 
Our contribution is the introduction of this refined evaluation metric, 
which offers a discriminative and comprehensive tool for assessing model performance. 
The HC-Scored metric facilitates a deeper understanding of LLM capabilities, 
guiding further advancements in the development of sophisticated language models.

\section{Results \& Analysis}\label{sec:analysis}
We present some selected interesting experimental results and analysis
for various scenarios in the following Section.
Full experimental results can be found in 
the Appendix~\ref{sec:full_experiments} due to the lack of space.

\subsection{Comparison of Java and Python}

Java and Python are both popular programming languages, 
each with its unique features and characteristics that influence
code structures and development practices.
One key difference between Java and Python is the syntax typing.
Java has to declare variables with their data types beforehand,
but Python variables can be assigned without explicit type declarations.
\begin{table}[!htb]
\centering
\small
\begin{tabular}{p{2.1cm}|p{0.3cm}p{0.3cm}p{0.45cm}|p{0.3cm}p{0.3cm}p{0.45cm}}
\toprule
{\bf Model} & \multicolumn{3}{c|}{\bf Java} & \multicolumn{3}{c}{\bf Python} \\
            & F1 & HC & HC$_{2}$ & F1 & HC & HC$_{2}$ \\\midrule
Decision Tree & 44.4 & 82.3 & 56.1 & 37.3 & 79.9 & 50.9 \\
Random Forest & 41.9 & 80.0 & 51.9 & 40.0 & 80.3 & 52.6 \\
SVM          & 24.3 & 71.9 & 39.1 & 17.2 & 66.8 & 36.0 \\\midrule
CodeBERT     & 77.3 & 90.9 & 80.2 & 73.3 & 88.7 & 76.5\\
GraphCodeBERT& 85.5 & 94.1 & 87.4 & 80.8 & 92.2 & 83.5\\
UniXcoder    & \bf{86.5} & \bf{94.6} & \bf{88.2} & \bf{85.4} & \bf{94.0} & \bf{87.4}\\\midrule
PLBART       & 85.3 & 94.3 & 87.1 & 77.2 & 91.0 & 80.7\\
CodeT5       & 82.4 & 93.1 & 84.8 & 75.5 & 89.9 & 79.1\\
CodeT5+      & 85.8 & \bf{94.6} & 88.0 & 78.2 & 91.0 & 81.2\\\midrule
Gemini Pro   & 29.7 & 80.1 & 48.6 & 33.4 & 80.2 & 51.8\\
ChatGPT 3.5  & 52.9 & 87.0 & 66.2 & 44.2 & 83.4 & 59.9\\
ChatGPT 4.0  & 60.6 & 90.0 & 72.7 & 52.7 & 87.2 & 67.4\\
\bottomrule
\end{tabular}
\caption{Performance comparison between Java and Python. Macro F1 score and HC scores are displayed.}
\label{tab:JavaPython}
\end{table}
UniXcoder was the best of the listed modes, and even the wrong answers lie in a
similar class which is hard to distinguish.
The HC scoring metric always boosts the score since it 
grants some score to wrong predictions. 
However, we can see by the HC score that LLMs are more affected than
the traditional and program language models. 
Since LLMs try to reason parts of the codes and combine those results into
a whole answer, LLMs tend to land closer to the correct answer if the parts are analyzed correctly.

\subsection{Effect of Code Length}
Intuitively, it is natural to assume that the shorter the code is, the easier it is to predict the complexity. To confirm our assumption, we categorize codes into four groups according to the number of tokens of the codes. If a code has less than or equal to 128 tokens, then the code falls into the first group (G1). If a code has more than 128 tokens and less than or equal to 256 tokens, then it falls into the second group (G2). The third group (G3) has codes with more than 256 tokens and less than or equal to 512 tokens. Lastly, group G4 has the longest codes, where each code has more than 512 tokens.

\begin{figure}[!htb]
\includegraphics[width=.97\linewidth, clip, trim={0.4cm 0.2cm 0.1cm 0.2cm}]{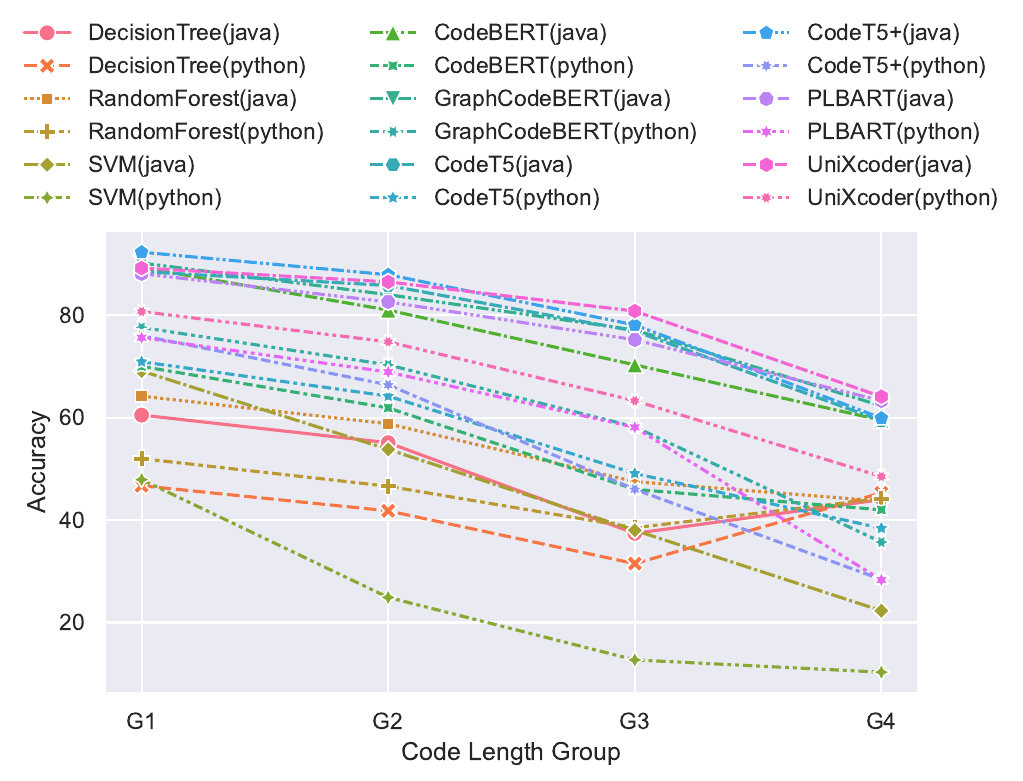}
\caption{Classification performance of models for different length groups of codes.}
\label{fig:length_java_and_python}
\end{figure}

Figure~\ref{fig:length_java_and_python} shows the experimental results on the four groups. It is easy to see that the experimental results confirm our assumption as the performance gets worse as the code becomes longer. We can see that the performance is quite similar in tendency except group G4. PLMs tend to have limits to their token size limiting their performance, however, the traditional methods rely on robust features of longer codes leading to better performance.

\subsection{Experiments with LLMs}
To predict the code time complexity, 
we use OpenAI’s ChatGPT (version 3.5 and 4.0) and Google's Gemini-Pro 
for closed-source LLMs\footnote{Accessed on 2024/2/12 for all closed-source LLMs} 
and use the instruction-tuned versions of Gemma (9b and 27b), Llama3.1 (8B and 70B),
Mistral-Nemo-2407 (12B), and Qwen2 (7B) for open-source LLMs. 
The prompt used for the experiments is given in Figure~\ref{fig:prompt}. 
The prompt specifies the role,
the question on time complexity, the output format,
and the answer choices. 
In summary of experimental results with LLMs, 
the biggest model, ChatGPT4.0 and Llama3.1 70B excel
for both the base and fine-tuned versions. 
Also, we can see that ChatGPT4.0 has a high score on HC$_2$, 
which indicates that ChatGPT4.0 was able to infer the time complexity to a point, 
but failed in gathering information to conclude the correct complexity.
\begin{figure}[htb]
\begin{tcolorbox}[colback=blue!5,colframe=blue!40!black,title=LLM response example, left=0.5mm,right=0.5mm,bottom=0.5mm,top=0.5mm,enlarge bottom by=-0.2cm]
"complexity": "linear", Here's why: the main logic of the code involves a binary search function, which has a time complexity of o(logn). However, the binary search is called only once, and the rest of the code runs in constant time. Therefore, the overall time complexity is dominated by the binary search, making it linear.
\end{tcolorbox}
\caption{Wrong LLM response example.}
\label{fig:response}
\end{figure}
We can see in Fig~\ref{fig:response}, LLMs sometimes fail on the final step of inference, 
which our HC scoring metric can capture. This is especially the case between Qwen2 and Qwen2.5, where Qwen2.5 is better on the F1 score but Qwen2 is better on the HC score.
Our experiments show that ChatGPT4.0 indeed has the best inference capabilities.
One thing to note is that some models, such as Gemma2 27B, 
are poor at analyzing time complexity.
This was due to the instruction-tuned models failing to output a response when
if it could not generate the desired output format. 
Gemma2 9B outputs the response ignoring the output format 
giving a better result than the 27B model. 
You can also see more results with smaller models in appendix~\ref{sec:full_experiments}.

\begin{table}[!tbh]
\centering
\begin{small}
\resizebox{\columnwidth}{!}{
\begin{tabular}{cr|ccccc}
\toprule
\multicolumn{2}{c|}{\bf Method} & Acc & F1 & HC & HC$_{2}$ & HC$_{3}$ \\
\midrule
\multicolumn{2}{r|}{Gemini Pro} & 34.0 & 31.6 & 80.2 & 50.2 & 63.1\\
\multicolumn{2}{r|}{ChatGPT 3.5} & 49.9 & 48.6 & 85.2 & 63.1 & 72.8\\
\multicolumn{2}{r|}{ChatGPT 4.0} & \bf{56.9} & \bf{56.7} & \bf{88.6} & \bf{70.1} & \bf{78.4}\\\midrule
\multirow{11}{*}{Inst}
 & CodeGemma-7B & 25.7 & 28.9 & 56.7 & 35.6 & 44.6\\
 & Gemma2-9B & 41.1 & 43.5 & 71.5 & 50.3 & 58.9\\
 & Gemma2-27B & 13.2 & 17.5 & 19.8 & 15.1 & 17.3\\
 & Gemma1.1-7B & 25.7 & 28.7 & 57.3 & 34.7 & 43.8\\
 & Llama3.2-3B& 22.9 & 22.8 & 60.6 & 33.1 & 43.4\\
 & Llama3.1-8B& 30.0 & 28.4 & 73.8 & 43.4 & 56.6\\
 & Llama3.1-70B& \bf{44.2} & 43.8 & \bf{81.3} & \bf{56.2} & \bf{67.1}\\
 & Mistral-12B& 42.3 & \bf{44.3} & 73.2 & 53.5 & 61.9\\
 & Qwen2.5-7B& 34.2 & 39.9 & 57.8 & 42.9 & 49.3\\
 & Qwen2.5-14B& 4.0 & 7.2 & 6.6 & 4.9 & 5.5\\
 & Qwen2-7B& 33.6 & 31.9 & 77.1 & 48.9 & 61.0\\\midrule
\multirow{11}{*}{\makecell{Fine-\\ tuned}}
 & CodeGemma-7B & 89.5 & 91.6 & 94.0 & 91.4 & 92.6\\
 & Gemma2-9B & 87.4 & 89.4 & 93.5 & 89.3 & 91.1\\
 & Gemma2-27B & 90.2 & 92.2 & 94.3 & 92.0 & 93.1\\
 & Gemma1.1-7B & 89.6 &91.6 & 94.0 & 91.5 & 92.7\\
 & Llama3.2-3B& 88.2 & 89.4 & 94.9 & 91.1 & 92.8\\
 & Llama3.1-8B& 89.4 & 90.7 & 95.4 & 92.0 & 93.6\\
 & Llama3.1-70B& \bf{92.9} & \bf{94.1} & \bf{96.0} & \bf{94.2} & \bf{95.0}\\
 & Mistral-12B& 88.5 & 89.7 & 94.8 & 90.9 & 92.7\\
 & Qwen2.5-7B& 88.7 & 90.0 & 95.0 & 91.4 & 93.1\\
 & Qwen2.5-14B& 91.9 & 93.2 & \bf{96.0} & 93.6 & 94.7\\
 & Qwen2-7B& 90.2 & 91.5 & 95.3 & 92.5 & 93.7\\
\bottomrule
\end{tabular}
}
\caption{Complexity prediction with accuracy, macro f1 score, and HC-Score for each LLM models}
\label{tab:LLMPickedexperiments}
\end{small}
\end{table}

\subsection{Qualitative Error Analysis}
After investigating the common errors from extensive experiments with many baseline models, we find that the following problems are the root causes of most error cases.

\paragraph{Unused Boilerplate Codes}
Codes can include parts of codes that are irrelevant to
the operation of the code. This can be because of coding habits
or template codes for handy development.
There are cases where the writer puts in ascii art in the comments.
These methods add to the overall recognition load of 
understanding the codebase and can obscure the true flow of execution.
Codes that include unused methods introduce noise, making it harder
for models to recognize the overall structure.

\paragraph{Logarithmic Loops}
The most common errors are from the logarithmic complexity class.
Loops with logarithmic sizes, such as those found in binary search algorithms, 
can significantly affect the prediction of a code's time complexity.
These loops have similar structures to normal linear loops,
but inside the loop, they have additional variables or conditions 
that control the algorithm's flow.
Unlike linear loops, this needs a thorough analysis of all contributing factors,
ensuring a comprehensive understanding of the algorithm's performance characteristics.
It seems like deep learning models lack the power to determine the contributing factors
and figure out their meaning and impact.

\paragraph{Too Much Conciseness of Python}

Despite the famous zen of Python
, it offers various ways to implement a loop such as classical for or while loop, list comprehension, and even lambda function. While the usage of list comprehension and lambda function makes Python codes much more concise and leads to statistics as in Figure~\ref{fig:features}, it also makes the complexity prediction task more challenging. The tendency is clearly seen especially when compared to Java in Table~\ref{tab:JavaPython}.



\section{Conclusions}
We have presented CodeComplex, the first large-scale bilingual benchmark dataset for predicting the worst-case time complexity of programs, 
designed to assess the reasoning abilities of LLMs in complex tasks. 
CodeComplex enhances LLM's ability for reasoning in complex time complexity prediction tasks by annotating a more extensive and diverse set of complexity labels and considering the input representation. We also provide a new scoring method to analyze the reasoning abilities of LLMs.
We have demonstrated the experimental results on several baseline models of classical to SoTA LLMs for benchmarking the complexity prediction performance. 
The results show that while the current state-of-the-art techniques provide promising baselines in assessing reasoning capabilities, 
there is still a long way to go to achieve a reliable performance for practical use cases.


\section{Limitations}
Although this study offers insights into enhancing LLM reasoning abilities 
in complex tasks such as code time complexity prediction tasks, 
future research should address the following limitations.
First, problem descriptions should be provided as part 
of the input for the completeness of the specification.
As shown in examples in Section~\ref{se:collection}, 
it is necessary to consider problem specification to determine 
the intended time complexity of the problem, 
which will apply to most of the solution codes for the problem. 
Second, we need to deal with the unintended biases towards 
problem-specific information rather than implementation details learned 
due to the characteristics of our dataset.
Finally, to fully address the inference capabilities of LLMs,
we could use methods such as chain of thought prompting 
or tree of thought prompting to handle cases such as Fig~\ref{fig:response}
and give us more robust results.

\bibliography{custom}

\onecolumn
\appendix

\section{Overview on CodeComplex Dataset}
\label{ssec:overview_dataset}

Our dataset construction process owes much to the recently released dataset called the CodeContests\footnote{\url{https://github.com/deepmind/code\_contests}}, a competitive programming dataset for machine learning by DeepMind.
We constructed a dataset with the codes from the CodeContests dataset that are again sourced from the coding competition platform Codeforces. Our dataset contains 4,120 codes in seven complexity classes, where there are new 500 Java source codes annotated with each complexity class. The seven complexity classes are constant ($O(1)$), linear ($O(n)$), quadratic ($O(n^2)$), cubic ($O(n^3)$), $O(\ln n)$, $O(n \ln n)$, and exponential. We also re-use 317 Java codes from CoRCoD as we confirmed that they also belong to the CodeContests dataset as the other 3803 codes during the dataset creation process.

For constructing the dataset, we asked twelve human annotators who have more than five years of programming experience and algorithmic expertise to inspect the codes manually and classify them into one of the seven complexity classes. Once each human annotator reported the initial result, we collected the annotation results and inspected them once again by assigning the initial result to two different annotators other than the initial annotator. Finally, we have collected 3803 complexity annotated codes in which there are 500 codes for each complexity class.

First, we selected several problems that are expected to belong to one of the considered complexity classes and submitted codes for the problems from Codeforces. The submitted codes contain both correct and incorrect solutions, and they are implemented in various programming languages such as C, C++, Java, and Python. We sorted out only the correct Java codes for our dataset construction. 

In the second step, before delving into the time complexity of problems, we divide the problems by the problem-solving strategy such as sorting, DP (dynamic programming), divide-and-conquer, DFS (depth-first search), BFS (breadth-first search), A*, and so on. This is because it is helpful to know the type of problem-solving strategy used to solve the problem for human annotators to analyze the time complexity, and problems solved by the same strategy tend to have similar time complexity. 

Third, we uniformly assign problems and correct codes for the problems to human annotators and let them carefully examine the problem-code pairs to label the time complexity of the codes. 
Notice that there can be solutions with different time complexities for a problem depending on how to actually implement the solutions.
We, therefore, provide a specific guideline that contains instructions and precautions to annotators so that human annotators can assign correct and consistent labels to the assigned codes. 

After the initial annotation process, we collect the results and assign them to different annotators to carefully cross-check the correctness of the initial annotation results. Primarily, we instruct the annotators again to carefully verify the results in accordance with the precautions provided in the annotation guideline.

\label{ssec:details_dataset}

\subsection{Further Details on CodeComplex Dataset Construction}
We gathered 128,000,000 submissions of Codeforces, where 4,086,507 codes are implemented in Java language. After discarding the incorrect codes (that do not pass all the test cases), there are 2,034,925 codes and 7,843 problems. Then the problems are split with their tags (e.g. sorting, dfs, greedy, etc) and given to the annotators with the guidelines in Section~\ref{ssec:guideline}. We were able to gather around 500 problems and 15,000 codes for the seven complexity classes.

As the complexity of codes for the same problem can vary depending on the implemented algorithms,
it is obvious that the codes we inspect also have various complexity classes.
However, we only target seven complexity classes that are the most frequently used complexity classes for algorithmic problems.
Accordingly, there were some codes we inspected which belong to other complexity classes such as $O(n^5)$ or $O(\ln{\ln{n}})$.
We inspected around 800 problems and found out that the complexity classes of approximately 15 of the problems belong outside the chosen complexity classes. Although it is still possible that one might implement codes with complexity class that falls into the seven complexity classes, we simply rule out the problems from our dataset to ease the annotation process.

During this process, we found out that many codes are not optimal for the given problem and some codes are too difficult to analyze due to their complex code structure. Moreover, there are many codes with a number of methods that are never used, mainly because the codes come from a coding competition platform and participants prefer just to include the methods that are frequently used in problem-solving regardless of the actual usage of the methods.

In the section below, we share the detailed guidelines provided to human annotators for a consistent and accurate annotation process.

\subsection{Guideline of Production}
\label{ssec:guideline}

\begin{tcolorbox}[enhanced,breakable,title=Annotator Guideline]
\begin{enumerate}[leftmargin=*]
    \item Check the variables that are described in the algorithm problems.
    Each algorithm implementation can have many variable instances
    and we only consider the variables that are given as inputs from the
    problems for calculating the time complexity.
    \item[*] For convenience,
    we use $n$ and $m$ in the guideline to denote the input variable and
    $|n|$ and $|m|$ to denote the size of $n$ and $m$.
    \item Considering the input variable from the prior step, follow the below instructions for each input type and
    calculate the time complexity.
    \begin{enumerate}[leftmargin=*]
        \item When only a number $n$ is given as an input,
        calculate the time complexity proportional to $n$.
        Do the same thing when there are two or more variables.
        For instance, when only $n$ is given as an input,
        the variable used to denote the time complexity of a code is
        $n$.
        \item When a number $n$ and $m$ numeric instances are given as
        inputs, calculate the time complexity proportional to the one with higher complexity.
        For instance, when $m=n^2$,
        we compute the complexity of a code
        with $m$. If the implemented algorithm runs in $O(n^2)=O(m)$,
        it belongs to the linear complexity class.
        \item If the input is given as
        constant values,
        the complexity of a given code
        also belongs to the constant class.
        For instance, if an algorithm problem states that exactly
        3 numeric values are given as
        inputs, the solution code only
        uses the constant number of operations.
        Therefore, the code belongs to the constant class.
    \end{enumerate}
    \item Consider the case where the code utilizes the input constraints of the problem.
    When the input is given by $n\le a$, the code can use the fixed value $a$ in the
    problem instead of using $n$. Mark these codes as unsuitable.
    \item Consider the built-in library that the implemented algorithm is using (e.g. HashMap, sort, etc.) to calculate the time complexity of an entire code.
    For instance, given $n$ numeric instances as inputs, when an implemented algorithm uses
    $O(n)$ iterations of built-in sort
    algorithm for $n$ numeric instances,
    the time complexity for the algorithm
    is $O(n^2\ln{n})$.
    \item When the code has unreachable codes, only consider the reachable code. 
    \item Mark the item that does
    not belong to any of the 7 complexity classes.
\end{enumerate}
\end{tcolorbox}

\newpage
\subsection{Statistics of CodeComplex}

Table~\ref{tab:enter-label} shows basic statistics in numbers of our CodeComplex dataset.
Each property is extracted using the abstract syntax tree of each language.

\begin{table*}[htb]
    \centering
    \setlength\tabcolsep{2.5pt}
    \begin{tabular}{l|rr|rr|rr|rr|rr|rr|rr|rr}
    \toprule
    {\bf Property} & \multicolumn{2}{c|}{$O(1)$} & \multicolumn{2}{c|}{$O(n)$} & \multicolumn{2}{c|}{$O(n^2)$} & \multicolumn{2}{c|}{$O(n^3)$} & \multicolumn{2}{c|}{$O(\ln{n})$} & \multicolumn{2}{c|}{$O(n\ln{n})$} & \multicolumn{2}{c|}{exponential} & \multicolumn{2}{c}{\bf Total}\\
    & Ja & Py &  Ja & Py &  Ja & Py &  Ja & Py &  Ja & Py &  Ja & Py &  Ja & Py &  Ja & Py  \\
    \midrule
    \#Problems & 38 & 50  & 94& 104  & 12&16  & 41&46  & 10&22  & 60&63  & 23&36 & 269&277 \\
    \#Lines & 31.7 & 19.7  & 60.9& 29.3  & 72.7&36.6  & 82.3&48.3  & 66.0&22.2  & 59.4&30.7  & 85.6&43.6 & 64.5&31.9  \\
    \#Functions & 2.6 & 1.3 & 4.5& 1.7 & 5.9&2.0 & 6.0&3.4 & 6.0&1.3 & 4.7&1.6 & 6.0&2.6 & 5.0&1.9 \\
    \#Variables & 5.3 & 9.7 & 12.2& 15.5 & 15.2&18.6 & 19.4&24.3 & 11.2&12.3 & 11.6&16.0 & 19.4&24.5 & 13.2&16.7 \\
    DoC & 5.7 & 1.5 & 10.2& 2.5 & 12.2&3.4 & 13.6&4.2 & 9.4&2.3 & 8.5&2.6 & 14.2&3.5 & 10.4&2.8 \\
    DoI & 0.6& 0.5 & 2.7& 1.0 & 4.0&1.0 & 5.5&1.0 & 1.8&0.8 & 2.4&0.9 & 5.6&0.9 & 3.1&0.9 \\
    \bottomrule
    \end{tabular}
    \caption{Statistics of codes from CodeComplex dataset. There are two values in each cell where the first value is about the Java codes and the second value is about the Python codes.}   
    \label{tab:enter-label}
\end{table*} 

\section{Experiment Details for each Model}

\paragraph{Hyperparameters}
\label{hyperparameters}
For all pre-trained programming language models, we use the AdamW~\cite{AdamW} optimizer 
with a warmup linear scheduler. The learning rate was set to 2e-6, 
epsilon to 1e-8, and the weight decay to 1e-2. 
We applied either the AutoTokenizer or the RobertaTokenizer.
The models were fine-tuned for 15 epochs before using them for evaluation.
For open-source LLMs, we used QLoRA and FSDP to train and test the models.
We used a AdamW optimizer with a warmup linear scheduler. The learning rate was set to
2e-5, and used flash attention for training. The LoRA alpha and r values were both set to 16,
with a dropout of 0.05. The quantization was set to 4 bits. We applied the basic 
AutoTokenizer and the basic chat template from the LLM model. Prompting was done in
a similar manner to Fig~\ref{fig:prompt}, where it includes system messages 
to give a coding expert role if the LLM supports a system role. Otherwise, it is
given as the header of the user message.

\begin{figure}[H]
\begin{tcolorbox}[colback=green!5,colframe=green!40!black,title=Prompt example,left=0.5mm,right=0.5mm,bottom=0.5mm,top=0.5mm,enlarge bottom by=-0.2cm]
You are a world expert in investigating properties of a code that influences the time complexity. 

The given code: "{[code]}"

Print "ONLY the time complexity in ONE WORD" of the given code in the answer from np, logn, quadratic, constant, cubic, linear and nlogn, do not print any other words in a json format.
\end{tcolorbox}
\caption{LLM prompt examples used in our experiments.}
\label{fig:prompt}
\end{figure}

\section{Full Experimental Results}
\label{sec:full_experiments}

\begin{table*}[!htb]
\centering
\begin{small}
\setlength\tabcolsep{4pt}
\renewcommand{\arraystretch}{0.8}
\begin{tabular}{l|ccccccc|cc}
\toprule
{\bf Method} & ${O(1)}$& ${O(\ln n)}$& ${O(n)}$& ${O(n\ln n)}$& ${O(n^2)}$& ${O(n^3)}$&   exponential & Micro & Macro\\
\midrule
Decision Tree & 64.4 & 55.9 & 15.2 & 65.2 & 68.8 & 32.9 & 34.4 & 48.6 & 48.1   \\
 Random Forest & 66.2 & 57.7 & 28.2 & 68.6 & 60.2 & 36.0 & 62.6 & 43.9 & 54.2  \\
SVM  & 49.7 & 40.0 & 65.1 & 42.7 & 74.6 & 23.5 & 18.0 & 28.1 & 44.8 \\
\midrule
 CodeBERT & 86.1 & 60.9 & 68.1 & 18.8 & 33.6 & 73.1 & 84.8 & 60.5 & 59.4  \\
GraphCodeBERT & 88.7 & 53.5 & 51.9 & 28.9 & 38.8 & 78.1 & 85.5 & 60.4 & 60.0  \\
{UniXCoder} & 83.1 & 54.8 & 54.1 & 15.9 & 33.9 & 76.9 & 88.1 & 57.7 & 56.6 \\
\midrule
 PLBART & 86.6 & 52.7 & 61.9 & 34.8 & 36.4 & 76.2 & 88.1 & 52.1 & 61.9 \\
 CodeT5  & 81.8 & 43.7 & 69.4 & 40.7 & 33.9 & 72.0 & 85.6 & \bf{60.7} & \bf{60.3} \\
CodeT5+  & 90.7 & 50.8 & 64.2 & 14.1 & 27.0 & 74.9 & 86.8 & 58.0 & 56.1  \\
\midrule
ChatGPT 3.5&54.0 &55.8 &74.1 &28.0 &67.7 &79.8 &85.73  &43.6 &35.6\\
ChatGPT 4.0&64.2 &43.4 &70.9 &72.8 &56.2 &67.2 &42.2 &54.8 &45.7\\
Gemini Pro&33.2 &15.4 &59.4 &7.1 &72.4 &8.1 &19.4 &30.1 &21.4\\
\bottomrule
\end{tabular}
\caption{Complexity prediction accuracy of classification methods for each complexity class on Java.}
\label{tab:class_java}
\end{small}
\end{table*}

\begin{table*}[!htb]
\centering
\begin{small}
\setlength\tabcolsep{4pt}
\renewcommand{\arraystretch}{0.8}
\begin{tabular}{l|ccccccc|cc}
\toprule
{\bf Method} & ${O(1)}$& ${O(\ln n)}$& ${O(n)}$& ${O(n\ln n)}$& ${O(n^2)}$& ${O(n^3)}$&   exponential & Micro & Macro\\
\midrule
Decision Tree & 45.0 & 39.8 & 37.0 & 62.4 & 42.1 & 65.8 & 6.6 & 38.8 & 42.7 \\
 Random Forest & 52.9 & 53.4 & 44.8 & 63.4 & 42.0 & 69.4 & 18.5 & 40.8 & 49.2  \\
SVM  & 43.1 & 25.3 & 78.6 & 52.1 & 14.0 & 20.7 & 13.7 & 23.6 & 35.4 \\
\midrule
CodeBERT & 68.0 & 66.1 & 31.7 & 46.9 & 40.6 & 63.8 & 25.6 &  51.2 & 49.2  \\
GraphCodeBERT & 68.5 & 56.9 & 61.9 & 51.4 & 56.8 & 68.1 & 34.8 & {\bf 58.1} & {\bf 57.3} \\
{UniXCoder} & 63.0 & 59.8 & 51.7 & 50.4 & 51.0 & 63.8 & 36.9 & 55.0 & 54.4 \\
\midrule
PLBART & 72.1 & 62.3 & 51.9 & 46.3 & 48.5 & 59.3 & 24.2 & 54.0 & 52.4 \\
CodeT5  & 68.9 & 47.1 & 44.5 & 41.4 & 43.6 & 51.8 & 38.3 & 48.9 & 48.4 \\
CodeT5+  & 65.9 & 54.9 & 58.9 & 23.4 & 40.3 & 66.3 & 24.6 & 49.8 & 47.7 \\
\midrule
ChatGPT 3.5 &44.4 &46.4 &83.0 &12.3 &60.6 &29.8 &19.6 &41.8 &35.6\\
ChatGPT 4.0 &54.7 &33.0 &80.0 &39.6 &61.4 &78.3 &34.2 &51.7 &41.9\\
Gemini Pro &35.9 &19.5 &72.0 &8.4 &61.3 &23.2 &16.6 &35.1 &28.5\\
\bottomrule
\end{tabular}
\caption{Complexity prediction accuracy of classification methods for each complexity class on Python.}
\label{tab:class_python}
\end{small}
\end{table*}

\begin{table*}[!htb]
\centering
\begin{small}
\setlength\tabcolsep{4pt}
\renewcommand{\arraystretch}{0.9}
\begin{tabular}{lr|ccccccc|cc}
\toprule
\multicolumn{2}{c|}{\bf Method} & ${O(1)}$& ${O(\ln n)}$& ${O(n)}$& ${O(n\ln n)}$& ${O(n^2)}$& ${O(n^3)}$&   exponential & Micro & Macro\\
\midrule
\multirow{14}{*}{Inst}
 & CodeGemma-7B & 85.9 & 85.5 & 76.5 & 72.6 & 83.5 & 86.8 & 90.2 & 25.7 & 24.4\\
 & Gemma2-9B & 86.7 &91.5 &78.5 &84.8 &81.9 &87.0 &82.6 &41.1 &36.1\\
 & Gemma2-27B & 81.0 &90.7 &84.8 &84.2 &85.8 &86.9 &90.3 &13.2 &13.6\\
 & Gemma1.1-2B & 79.9 & 89.2 & 74.3 & 84.1 & 60.1 & 86.9 & 61.8 & 12.8 & 9.1\\
 & Gemma1.1-7B & 83.7 & 90.6 & 76.9 & 76.1 & 82.1 & 85.6 & 80.7 & 25.7 & 23.3\\
 & Llama3.1-8B& 83.7 &90.3 &58.5 &78.4 &79.1 &87.9 &89.7 &30.0 &23.4\\
 & Llama3.1-70B& 87.3 &90.4 &69.7 &75.5 &86.1 &90.0 &92.2 &\bf{44.2} &\bf{36.6}\\
 & Llama3.2-1B & 79.2 & 89.4 & 78.8 & 82.4 & 83.6 & 86.7 & 75.1 & 8.2 & 9.2\\
 & Llama3.2-3B & 84.4 & 89.5 & 78.4 & 82.9 & 56.7 & 87.0 & 82.2 & 22.9 & 17.8\\
 & Mistral-12B& 88.6 &91.1 &76.1 &88.5 &78.3 &88.6 &83.0 &42.3 &\bf{36.6}\\
 & Qwen2.5-1.5B& 80.0 & 90.0 & 82.4 & 84.1 & 86.3 & 86.8 & 85.3 & 1.4 & 2.2\\
 & Qwen2.5-7B& 85.3 & 91.9 & 79.3 & 84.6 & 80.9 & 87.1 & 91.5 & 34.2 & 34.1\\
 & Qwen2.5-14B& 80.2 & 90.6 & 82.7 & 84.4 & 85.7 & 86.7 & 89.2 & 4.0 & 6.5\\
 & Qwen2-7B& 89.1 &91.3 &76.0 &84.1 &51.9 &88.2 &90.9 &33.6 &25.7\\\midrule
\multirow{14}{*}{\makecell{Fine-\\ tuned}}
 & CodeGemma-7B & 97.9 & 97.7 & 96.2 & 96.8 & 97.8 & 98.4 & 98.8 & 89.5 & 80.1\\
 & Gemma2-9B & 97.0 &97.7 &95.0 &96.8 &96.5 &98.1 &98.2 &87.4 &78.2\\
 & Gemma2-27B & 97.8 &98.2 &96.3 &97.1 &98.1 &98.3 &99.1 &90.2 &80.8\\
 & Gemma1.1-2B & 96.9 & 97.3 & 94.6 & 94.5 & 94.6 & 97.2 & 98.4 & 84.6 & 75.7\\
 & Gemma1.1-7B & 98.3 & 98.1 & 96.1 & 96.9 & 97.3 & 98.2 & 98.8 & 89.6 & 80.1\\
 & Llama3.1-8B& 98.1 &98.4 &95.1 &96.1 &96.8 &98.5 &98.6 &89.4 &79.4\\
 & Llama3.1-70B& 98.4 &99.0 &97.6 &97.9 &97.8 &99.0 &99.0 &\bf{92.9} &\bf{82.4}\\
 & Llama3.2-1B & 96.4 & 97.6 & 91.4 & 94.5 & 95.1 & 97.9 & 97.9 & 83.9 & 74.8\\
 & Llama3.2-3B & 97.4 & 97.9 & 93.7 & 95.8 & 97.3 & 98.5 & 98.5 & 88.2 & 78.5\\
 & Mistral-12B& 97.6 &98.3 &93.7 &96.1 &96.9 &98.6 &98.7 &88.5 &78.8\\
 & Qwen2.5-1.5B& 98.2 & 97.9 & 94.8 & 95.3 & 96.5 & 98.6 & 98.7 & 88.6 & 78.7\\
 & Qwen2.5-7B& 98.3 & 98.2 & 94.8 & 95.3 & 96.3 & 98.9 & 98.7 & 88.7 & 78.9\\
 & Qwen2.5-14B& 98.5 & 98.2 & 96.6 & 97.8 & 97.9 & 99.1 & 98.7 & 91.9 & 81.5\\
 & Qwen2-7B& 98.1 &98.0 &95.4 &96.3 &98.1 &98.7 &98.7 &90.2 &80.1\\
\bottomrule
\end{tabular}
\caption{Complexity prediction accuracy of classification methods for each complexity class on open source LLMs.}
\label{tab:class_LLM}
\end{small}
\end{table*}

\subsection{Performance Per Complexity Class}
We analyzed the results with responses that could be recognized as one of the seven classes.
We can see in Table~\ref{tab:class_java}, Table~\ref{tab:class_python} and Table~\ref{tab:LLMFullexperiments}, 
that the constant class and linear class are the most accurate among the classes.
The direct correlation between the input size and the number 
of iterations makes these classes easy to analyze.
However, models still face challenges in predicting the complexity of self-referential algorithms
and control flows that are not in the main part of the algorithm, such as predicting between
the $O(n\log{n})$ class and $O(n^2)$ class. Failing to analyze loop controls leads to failing to
differentiate the $O(n\log n)$ case from $O(n^2)$. 
This is well seen in the confusion matrix in Section~\ref{sec:conf}, where we can see many
wrong predictions are between classes $O(n)$, $O(n\log{n})$, and $O(n^2)$.
\begin{table}[!htb] 
    \centering
    \begin{tabular}{l|cccc|cccc}
    \toprule
    {\bf Method} & {\bf G1} &  {\bf G2} &  {\bf G3} & {\bf G4} & {\bf G1} &  {\bf G2} &  {\bf G3} & {\bf G4}  \\
    & \multicolumn{4}{c|}{Java} & \multicolumn{4}{c}{Python} \\
    \midrule
    Decision Tree & 57.2&45.6 & 40.0 & 38.2  & 57.2&45.6 & 40.0 & 38.2   \\
    Random Forest & 62.3& 46.8 & 40.6 & 26.4 & 62.3& 46.8 & 40.6 & 26.4   \\
    SVM  & 48.9 & 18.1 & 18.1 & 16.6 & 48.9 & 18.1 & 18.1 & 16.6  \\
    \midrule
    CodeBERT& 72.4 & 62.8 & 60.7 &48.0 & 56.9 & 46.9 & 37.5 &22.8 \\
    GraphCodeBERT& 74.6 & 61.7 & 49.8 &39.4 & 60.3 & 57.8 & 44.1 &30.8 \\
    UniXCoder & 58.6 & 54.4 & 43.2 & 31.2 & 58.6 & 54.4 & 43.2 & 31.2\\
    \midrule
    PLBART & 74.3 &  65.1 & 62.5 &52.8 & 60.6 &  49.4 & 39.9 &23.2 \\
    CodeT5& 69.5 & 56.5 & 52.4 & 42.4 & 53.6 & 48.1 & 36.5 & 19.5\\
    CodeT5+& 72.8 & 63.5 & 53.0 & 44.4 & 56.4 &  42.4 &  30.7 &  29.8\\
    \bottomrule
    \end{tabular}
    \caption{Prediction performance on different code length groups.}
    \label{tab:length_java}
\end{table}

\subsection{Does Problem Description Help?}
\begin{table}[!htb]
\centering
\begin{tabular}{l|cc|cc}\toprule
{\bf Model} & \multicolumn{2}{c|}{\bf w/o Desc.} & \multicolumn{2}{c}{\bf with Desc.} \\
            & Ja & Py & Ja & Py \\
\midrule
ChatGPT 3.5     & {\bf 43.38} &  {\bf 43.14}   & 42.51 &  36.55\\
ChatGPT 4.0     & 55.42  &  51.57   & {\bf 57.61}  & {\bf 54.28} \\
\bottomrule
\end{tabular}
\caption{Complexity prediction performances of LLMs with and without a problem description in the prompt by the help of information retrieval.}
\label{tab:spec_help}
\end{table}
A critical aspect in accurately determining the worst-case time complexity of a given code is the comprehensive understanding of the problem specifications. In certain instances, these specifications may indicate that some inputs are constant, significantly influencing the complexity analysis. The absence of a full and detailed specification can lead to an incomplete or incorrect assessment of the worst-case time complexity. 
Table~\ref{tab:spec_help} shows that problem descriptions actually help LLMs perform better as ChatGPT 4.0 is known to have real-time access to the information on the web. Note that the performance becomes worse for ChatGPT 3.5 when problem IDs are provided, as ChatGPT 3.5 does not utilize the problem descriptions, only from problem IDs.

\subsection{Effect of Dead Code Elimination}
\begin{table}[!htb]
\centering
\renewcommand{\arraystretch}{0.8}
\begin{tabular}{l|cc|cc}\toprule
{\bf Model} & \multicolumn{2}{c|}{\bf After} & \multicolumn{2}{c}{\bf Before} \\
            & Ja & Py & Ja & Py \\
\midrule
Decision Tree & {\bf 48.6} & {\bf 38.8} & 47.6 & 21.1  \\
Random Forest & {\bf 43.9} & {\bf 40.8} & 43.2 & 23.0 \\
SVM          & {\bf 28.1} & {\bf 23.6} & 27.1 & 21.5 \\\midrule
CodeBERT     & {\bf 60.5} & 51.2 & 59.7 & {\bf 52.0} \\
GraphCodeBERT& {\bf 60.4} & 58.1 & 57.8 & {\bf 60.2} \\
UniXcoder    & {\bf 57.7} & 55.0 & 57.2 & {\bf 55.3} \\\midrule
PLBART       & {\bf 62.1} & 54.0 & 61.2 & {\bf 55.4} \\
CodeT5       & {\bf 60.7} & 48.9 & 60.2 & {\bf 49.8} \\
CodeT5+      & {\bf 58.0} & 49.8 & 57.4 & {\bf 50.0} \\
\midrule
ChatGPT 3.5& {\bf 54.1} & 45.4 &53.8 & {\bf 46.16} \\
ChatGPT 4.0 & 59.9 & {\bf 53.7} &{\bf 59.7} &53.8 \\
Gemini Pro & 32.4 & {\bf 35.1} & {\bf 34.0}&33.5 \\
\bottomrule
\end{tabular}
\caption{Performance comparison before and after dead code processing.}
\label{tab:Origin2Dead}
\end{table}

Table~\ref{tab:Origin2Dead} shows the effect of dead code elimination on prediction performance.
Given the nature of codes submitted to competitive programming platforms, there are a lot of redundant variables, methods and even classes in the codes.
Due to Java's complicated IO functions and limited built-in data structures, there are many codes
related to the implementation of IO and data structures. 
Removing such fragments helps the models concentrate
on the program structure and results in enhanced prediction accuracy.
On the other hand, it appears that the dead code elimination does not help improve the performance on Python as the Python codes are already more concise than the Java codes due to its own language design principle.





\begin{table}[tbh]
\centering
\begin{tabular}{cr|ccccc}
\toprule
\multicolumn{2}{c|}{\bf Method} & Acc & F1 & HC & HC$_{2}$ & HC$_{3}$ \\
\midrule
\multicolumn{2}{r|}{Gemini Pro} & 34.0 & 31.6 & 80.2 & 50.2 & 63.1\\
\multicolumn{2}{r|}{ChatGPT 3.5} & 49.9 & 48.6 & 85.2 & 63.1 & 72.8\\
\multicolumn{2}{r|}{ChatGPT 4.0} & \bf{56.9} & \bf{56.7} & \bf{88.6} & \bf{70.1} & \bf{78.4}\\\midrule
\multirow{14}{*}{Inst}
 & CodeGemma-7B & 25.7 & 28.9 & 56.7 & 35.6 & 44.6\\
 & Gemma2-9B & 41.1 & 43.5 & 71.5 & 50.3 & 58.9\\
 & Gemma2-27B & 13.2 & 17.5 & 19.8 & 15.1 & 17.3\\
 & Gemma1.1-2B & 12.8 & 10.5 & 53.5 & 22.6 & 32.7\\
 & Gemma1.1-7B & 25.7 & 28.7 & 57.3 & 34.7 & 43.8\\
 & Llama3.2-1B& 8.2 & 11.1 & 24.4 & 12.0 & 15.9\\
 & Llama3.2-3B& 22.9 & 22.8 & 60.6 & 33.1 & 43.4\\
 & Llama3.1-8B& 30.0 & 28.4 & 73.8 & 43.4 & 56.6\\
 & Llama3.1-70B& \bf{44.2} & 43.8 & \bf{81.3} & \bf{56.2} & \bf{67.1}\\
 & Mistral-12B& 42.3 & \bf{44.3} & 73.2 & 53.5 & 61.9\\
 & Qwen2.5-1.5B& 1.4 & 2.0 & 4.6 & 2.2 & 2.9\\
 & Qwen2.5-7B& 34.2 & 39.9 & 57.8 & 42.9 & 49.3\\
 & Qwen2.5-14B& 4.0 & 7.2 & 6.6 & 4.9 & 5.5\\
 & Qwen2-7B& 33.6 & 31.9 & 77.1 & 48.9 & 61.0\\\midrule
\multirow{14}{*}{\makecell{Fine-\\ tuned}}
 & CodeGemma-7B & 89.5 & 91.6 & 94.0 & 91.4 & 92.6\\
 & Gemma2-9B & 87.4 & 89.4 & 93.5 & 89.3 & 91.1\\
 & Gemma2-27B & 90.2 & 92.2 & 94.3 & 92.0 & 93.1\\
 & Gemma1.1-2B & 84.6 & 86.5 & 92.7 & 88.0 & 90.0\\
 & Gemma1.1-7B & 89.6 &91.6 & 94.0 & 91.5 & 92.7\\
 & Llama3.2-1B& 83.9 & 85.2 & 93.2 & 87.1 & 89.8\\
 & Llama3.2-3B& 88.2 & 89.4 & 94.9 & 91.1 & 92.8\\
 & Llama3.1-8B& 89.4 & 90.7 & 95.4 & 92.0 & 93.6\\
 & Llama3.1-70B& \bf{92.9} & \bf{94.1} & \bf{96.0} & \bf{94.2} & \bf{95.0}\\
 & Mistral-12B& 88.5 & 89.7 & 94.8 & 90.9 & 92.7\\
 & Qwen2.5-1.5B& 88.6 & 89.8 & 95.2 & 91.6 & 93.3\\
 & Qwen2.5-7B& 88.7 & 90.0 & 95.0 & 91.4 & 93.1\\
 & Qwen2.5-14B& 91.9 & 93.2 & \bf{96.0} & 93.6 & 94.7\\
 & Qwen2-7B& 90.2 & 91.5 & 95.3 & 92.5 & 93.7\\
\bottomrule
\end{tabular}
\caption{Complexity prediction with accuracy, macro f1 score, and HC-Score for each LLM models}
\label{tab:LLMFullexperiments}
\end{table}

\section{Failure Cases}

\begin{javacode}
public class mad {
    public static void main(String[] args) {
        Scanner sc = new Scanner(System.in);
        int cura = 0, curb = 0;
        int ver;
        System.out.println("? 0 0");
        System.out.flush();
        ver = sc.nextInt();
        for (int i = 29; i >= 0; i--) {
            System.out.println("? " + (cura + (1 << i)) + " " + curb);
            System.out.flush();
            int temp1 = sc.nextInt();
            System.out.println("? " + cura + " " + (curb + (1 << i)));
            System.out.flush();
            int temp2 = sc.nextInt();
            if (temp1 != temp2) {
                if (temp2 == 1) {
                    cura += (1 << i);
                    curb += (1 << i);
                }
            } else {
                if (ver == 1) cura += (1 << i);
                if (ver == -1) curb += (1 << i);
                ver = temp1;
            }
        }
        System.out.println("! " + cura + " " + curb);
    }
}
\end{javacode}
The following example exhibits a failure example where our model predicts $O(2^n)$ for a code with $O(\ln n)$ complexity. We suspect that the primary reason is the usage of bitwise operators. When we filter the codes that use any bitwise operator at least once from our CodeComplex dataset, about 56 of the codes belong to the class $O(2^n)$. We find that many implementations for exponential problems rely on bitwise operators as they can efficiently manage the backtracking process by manipulating bit-level flags.

The following example demonstrates the case when our model predicts constant time complexity $O(1)$ for a code that runs in $O(n)$ time. We suspect that our model may have ignored the existence of the \texttt{check} method which actually determines the $O(n)$ time complexity or considered the argument of \texttt{check} as constant.
\begin{javacode}
public class abc {
    public static int check(StringBuilder s) {
        int countRemove = 0;
        if (!s.toString().contains("xxx")) return countRemove;
        else {
            for (int i = 1; i < s.length() - 1; i++) {
                if (s.charAt(i - 1) == 'x' && s.charAt(i) == 'x' && s.charAt(i + 1) == 'x') {
                    countRemove++;
                }
            }
            return countRemove;
        }
    }

    public static void main(String[] args) {
        Scanner sc = new Scanner(System.in);
        int n = sc.nextInt();
        String s = sc.next();
        StringBuilder sb = new StringBuilder("");
        sb.append(s);
        System.out.println(check(sb));
    }
}
\end{javacode}

The following is the case where our model predicts the quadratic time complexity $O(n^2)$ when the ground-truth label is $O(n \ln n)$. We guess that our model simply translates the nested \texttt{for} loops into the quadratic time complexity. However, the outer loop is to repeat each test case and therefore should be ignored. Then, the $O(n \ln n)$ complexity can be determined by the \texttt{sort} function used right before the nested loops.
\begin{javacode}
public class round111A {
    public static void main(String[] args) {
        Scanner sc = new Scanner(System.in);
        int n = sc.nextInt();
        int[] coins = new int[n];
        for (int i = 0; i < n; ++i) coins[i] = sc.nextInt();
        Arrays.sort(coins);
        int ans = (int) 1e9;
        for (int i = 1; i <= n; ++i) {
            int sum1 = 0;
            int c = 0;
            int j = n - 1;
            for (j = n - 1; j >= 0 && c < i; --j, ++c) {
                sum1 += coins[j];
            }
            int sum2 = 0;
            for (int k = 0; k <= j; ++k) sum2 += coins[k];
            if (sum1 > sum2) {
                System.out.println(i);
                return;
            }
        }
    }
}
\end{javacode}

The following is the case when our model is confused with exponential complexity $O(2^n)$ with quadratic complexity $O(n^2)$. The code actually runs in exponential time in the worst-case but our model simply returns quadratic time complexity as it does not take into account the recursive nature of the method \texttt{solve}.

\begin{javacode}
public class D {
    static int n, KA, A;
    static int[] b;
    static int[] l;
    static double ans = 0;

    public static void main(String[] args) throws IOException {
        Scanner in = new Scanner(System.in);
        n = in.nextInt();
        KA = in.nextInt();
        A = in.nextInt();
        b = new int[n];
        l = new int[n];
        for (int i = 0; i < l.length; i++) {
            b[i] = in.nextInt();
            l[i] = in.nextInt();
        }
        dp = new double[n + 2][n + 2][n * 9999 + 2];
        go(0, KA);
        System.out.printf("%.6f\n", ans);
    }

    public static void go(int at, int k) {
        if (at == n) {
            ans = Math.max(ans, solve(0, 0, 0));
            return;
        }
        for (int i = 0; i <= k; i++) {
            if (l[at] + i * 10 <= 100) {
                l[at] += i * 10;
                go(at + 1, k - i);
                l[at] -= i * 10;
            }
        }
    }

    static double dp[][][];

    public static double solve(int at, int ok, int B) {
        if (at == n) {
            if (ok > n / 2) {
                return 1;
            } else {
                return (A * 1.0) / (A * 1.0 + B);
            }
        }
        double ret = ((l[at]) / 100.0) * solve(at + 1, ok + 1, B) + (1.0 - ((l[at]) / 100.0)) * solve(at + 1, ok, B + b[at]);
        return ret;
    }
    
}
\end{javacode}

The following is the case when our model predicts $O(\ln n)$ for a code with $O(n^2)$ complexity. It is easily seen that the \texttt{inversions} function determines the quadratic time complexity by the nested \texttt{for} loops. We suspect that somehow our model does not take into account the \texttt{inversions} function in the complexity prediction and instead focuses on the modulo () operator to draw the wrong conclusion that the complexity is in $O(\ln n)$.

\begin{javacode}
public class maestro {
    public static long inversions(long[] arr) {
        long x = 0;
        int n = arr.length;
        for (int i = n - 2; i >= 0; i--) {
            for (int j = i + 1; j < n; j++) {
                if (arr[i] > arr[j]) {
                    x++;
                }
            }
        }
        return x;
    }

    public static void main(String[] args) {
        Scanner sc = new Scanner(System.in);
        int n = sc.nextInt();
        long[] arr = new long[n];
        for (int i = 0; i < n; i++) arr[i] = sc.nextLong();
        long m = sc.nextLong();
        long x = inversions(arr) % 2;
        for (int i = 0; i < m; i++) {
            int l = sc.nextInt() - 1;
            int r = sc.nextInt() - 1;
            if ((r - l + 1) % 4 > 1) x = (x + 1) % 2;
            if (x == 1) System.out.println("odd");
            else System.out.println("even");
        }
    }
}
\end{javacode}

\section{Further Details on Dead Code Elimination}


In a broad sense, the dead code includes redundant code, unreachable code, oxbow code, and so on. We only focus on eliminating unreachable codes, mainly methods and classes that are declared but used nowhere in the code. In order to find such dead codes, we first parse a Java code into an AST and discover methods and classes that do not exist in any method call, class declaration, and arguments of methods. Once we discover such unused methods and classes, we remove the codes corresponding to the declarations of these methods and classes.

The following codes are a running example of the dead code elimination process. From the first code, we can obtain the second code by applying the dead code elimination. It is readily seen that the number of lines decreases from 211 to 101 by the elimination process. In fact, our model predicts $O(\ln n)$ and $O(1)$ for the complexity of the code before and after dead code elimination, respectively, while the actual complexity of the code is $O(1)$.

\begin{javacode}
public class Main {
    static long mod = ((long) 1e9) + 7;

    public static int gcd(int a, int b) {
        if (b == 0) return a;
        else return gcd(b, a % b);
    }

    public static long pow_mod(long x, long y) {
        long res = 1;
        x = x % mod;
        while (y > 0) {
            if ((y & 1) == 1) res = (res * x) % mod;
            y = y >> 1;
            x = (x * x) % mod;
        }
        return res;
    }

    public static int lower_bound(int[] arr, int val) {
        int lo = 0;
        int hi = arr.length - 1;
        while (lo < hi) {
            int mid = lo + ((hi - lo + 1) / 2);
            if (arr[mid] == val) {
                return mid;
            } else if (arr[mid] > val) {
                hi = mid - 1;
            } else lo = mid;
        }
        if (arr[lo] <= val) return lo;
        else return -1;
    }

    public static int upper_bound(int[] arr, int val) {
        int lo = 0;
        int hi = arr.length - 1;
        while (lo < hi) {
            int mid = lo + ((hi - lo) / 2);
            if (arr[mid] == val) {
                return mid;
            } else if (arr[mid] > val) {
                hi = mid;
                ;
            } else lo = mid + 1;
        }
        if (arr[lo] >= val) return lo;
        else return -1;
    }

    public static void main(String[] args) throws java.lang.Exception {
        Reader sn = new Reader();
        Print p = new Print();
        int n = sn.nextInt();
        while ((n--) > 0) {
            int a = sn.nextInt();
            int b = sn.nextInt();
            int small = Math.min(a, b);
            int large = Math.max(a, b);
            long steps = 0;
            while (small != 0) {
                steps += (long) (large / small);
                int large1 = small;
                small = large % small;
                large = large1;
            }
            p.printLine(Long.toString(steps));
        }
        p.close();
    }
}

class Pair implements Comparable<Pair> {
    int val;
    int in;

    Pair(int a, int b) {
        val = a;
        in = b;
    }

    @Override
    public int compareTo(Pair o) {
        if (val == o.val) return Integer.compare(in, o.in);
        else return Integer.compare(val, o.val);
    }
}

class Reader {
    final private int BUFFER_SIZE = 1 << 16;
    private DataInputStream din;
    private byte[] buffer;
    private int bufferPointer, bytesRead;

    public boolean isSpaceChar(int c) {
        return c == ' ' || c == '\n' || c == '\r' || c == '\t' || c == -1;
    }

    public Reader() {
        din = new DataInputStream(System.in);
        buffer = new byte[BUFFER_SIZE];
        bufferPointer = bytesRead = 0;
    }

    public Reader(String file_name) throws IOException {
        din = new DataInputStream(new FileInputStream(file_name));
        buffer = new byte[BUFFER_SIZE];
        bufferPointer = bytesRead = 0;
    }

    public String readLine() throws IOException {
        byte[] buf = new byte[64];
        int cnt = 0, c;
        while ((c = read()) != -1) {
            if (c == '\n') break;
            buf[cnt++] = (byte) c;
        }
        return new String(buf, 0, cnt);
    }

    public String readWord() throws IOException {
        int c = read();
        while (isSpaceChar(c)) c = read();
        StringBuilder res = new StringBuilder();
        do {
            res.appendCodePoint(c);
            c = read();
        } while (!isSpaceChar(c));
        return res.toString();
    }

    public int nextInt() throws IOException {
        int ret = 0;
        byte c = read();
        while (c <= ' ') c = read();
        boolean neg = (c == '-');
        if (neg) c = read();
        do {
            ret = ret * 10 + c - '0';
        } while ((c = read()) >= '0' && c <= '9');
        if (neg) return -ret;
        return ret;
    }

    public long nextLong() throws IOException {
        long ret = 0;
        byte c = read();
        while (c <= ' ') c = read();
        boolean neg = (c == '-');
        if (neg) c = read();
        do {
            ret = ret * 10 + c - '0';
        } while ((c = read()) >= '0' && c <= '9');
        if (neg) return -ret;
        return ret;
    }

    public double nextDouble() throws IOException {
        double ret = 0, div = 1;
        byte c = read();
        while (c <= ' ') c = read();
        boolean neg = (c == '-');
        if (neg) c = read();
        do {
            ret = ret * 10 + c - '0';
        } while ((c = read()) >= '0' && c <= '9');
        if (c == '.') {
            while ((c = read()) >= '0' && c <= '9') {
                ret += (c - '0') / (div *= 10);
            }
        }
        if (neg) return -ret;
        return ret;
    }

    private void fillBuffer() throws IOException {
        bytesRead = din.read(buffer, bufferPointer = 0, BUFFER_SIZE);
        if (bytesRead == -1) buffer[0] = -1;
    }

    private byte read() throws IOException {
        if (bufferPointer == bytesRead) fillBuffer();
        return buffer[bufferPointer++];
    }

    public void close() throws IOException {
        if (din == null) return;
        din.close();
    }
}

class Print {
    private final BufferedWriter bw;

    public Print() {
        bw = new BufferedWriter(new OutputStreamWriter(System.out));
    }

    public void print(String str) throws IOException {
        bw.append(str);
    }

    public void printLine(String str) throws IOException {
        print(str);
        bw.append("\n");
    }

    public void close() throws IOException {
        bw.close();
    }
}
\end{javacode}

Code after Dead Code Elimination:
\begin{javacode}
    static long mod = ((long) 1e9 + 7);

    public static int gcd(int a, int b) {
        if ((b == 0)) return a;
        else return gcd(b, (a % b));
    }

    public static void main(String[] args) throws java.lang.Exception {
        Reader sn = new Reader();
        Print p = new Print();
        int n = sn.nextInt();
        while ((n > 0)) {
            int a = sn.nextInt();
            int b = sn.nextInt();
            int small = Math.min(a, b);
            int large = Math.max(a, b);
            long steps = 0;
            while ((small != 0)) {
                steps += (long) (large / small);
                int large1 = small;
                small = (large % small);
                large = large1;
            }
            p.printLine(Long.toString(steps));
        }
        p.close();
    }

class Reader {
    final private int BUFFER_SIZE = (1 << 16);
    private DataInputStream din;
    private byte[] buffer;
    private int bufferPointer, bytesRead;

    public boolean isSpaceChar(int c) {
        return (((((c == ' ') || (c == '\n')) || (c == '\r')) || (c == '\t')) || (c == -1));
    }

    public Reader() {
        din = new DataInputStream(System.in);
        buffer = new byte[BUFFER_SIZE];
        bufferPointer = bytesRead = 0;
    }

    public Reader(String file_name) throws IOException {
        din = new DataInputStream(new FileInputStream(file_name));
        buffer = new byte[BUFFER_SIZE];
        bufferPointer = bytesRead = 0;
    }

    public int nextInt() throws IOException {
        int ret = 0;
        byte c = read();
        while ((c <= ' ')) c = read();
        boolean neg = (c == '-');
        if (neg) c = read();
        do {
            ret = (((ret * 10) + c) - '0');
        } while ((((c = read()) >= '0') && (c <= '9')));
        if (neg) return -ret;
        return ret;
    }

    private void fillBuffer() throws IOException {
        bytesRead = din.read(buffer, bufferPointer = 0, BUFFER_SIZE);
        if ((bytesRead == -1)) buffer[0] = -1;
    }

    private byte read() throws IOException {
        if ((bufferPointer == bytesRead)) fillBuffer();
        return buffer[bufferPointer++];
    }

    public void close() throws IOException {
        if ((din == null)) return;
        din.close();
    }
}

class Print {
    final private BufferedWriter bw;

    public Print() {
        bw = new BufferedWriter(new OutputStreamWriter(System.out));
    }

    public void print(String str) throws IOException {
        bw.append(str);
    }

    public void printLine(String str) throws IOException {
        print(str);
        bw.append("\n");
    }
    public void close() throws IOException {
        bw.close();
    }
}
\end{javacode}

\newpage
\section{Confusion Matrices for Models}
\label{sec:conf}

\begin{figure*}[!htb]
    \centering
    \includegraphics[width=0.47\textwidth]{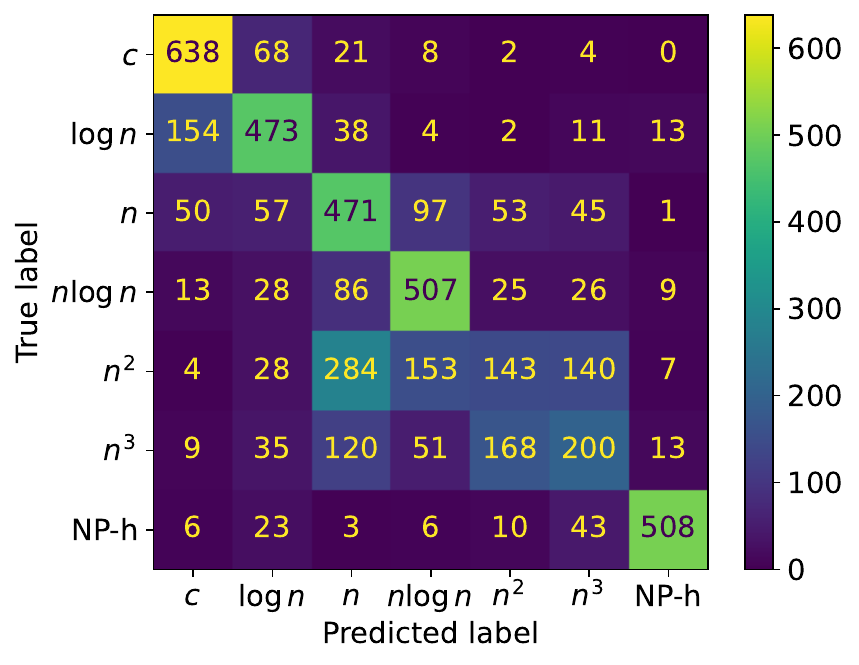}
    \includegraphics[width=0.47\textwidth]{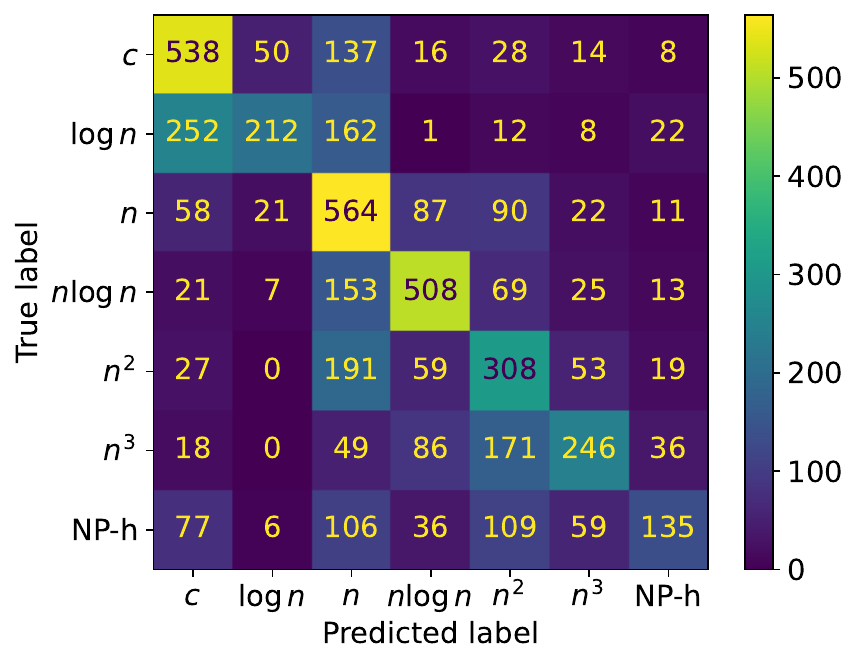}
    \caption{Confusion matrices of prediction results by CodeBERT on Java (left) and Python (right) datasets.}
    \label{fig:conf_CodeBERT}
\end{figure*}

\begin{figure*}[!htb]
    \centering
    \includegraphics[width=0.47\textwidth]{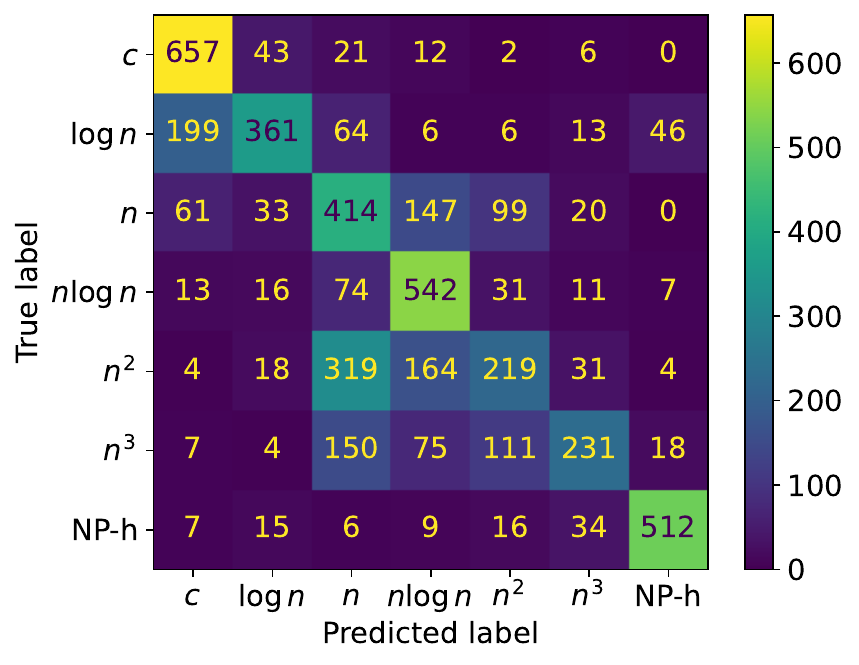}
    \includegraphics[width=0.47\textwidth]{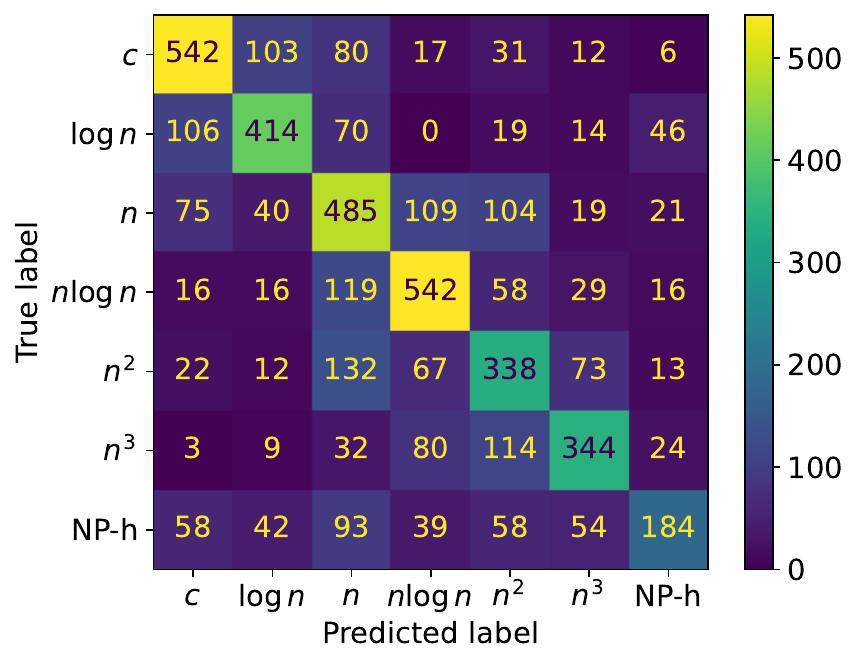}
    \caption{Confusion matrices of prediction results by GraphCodeBERT on Java (left) and Python (right) datasets.}
    \label{fig:conf_GraphCodeBERT}
\end{figure*}

\begin{figure*}[!htb]
    \centering
    \includegraphics[width=0.47\textwidth]{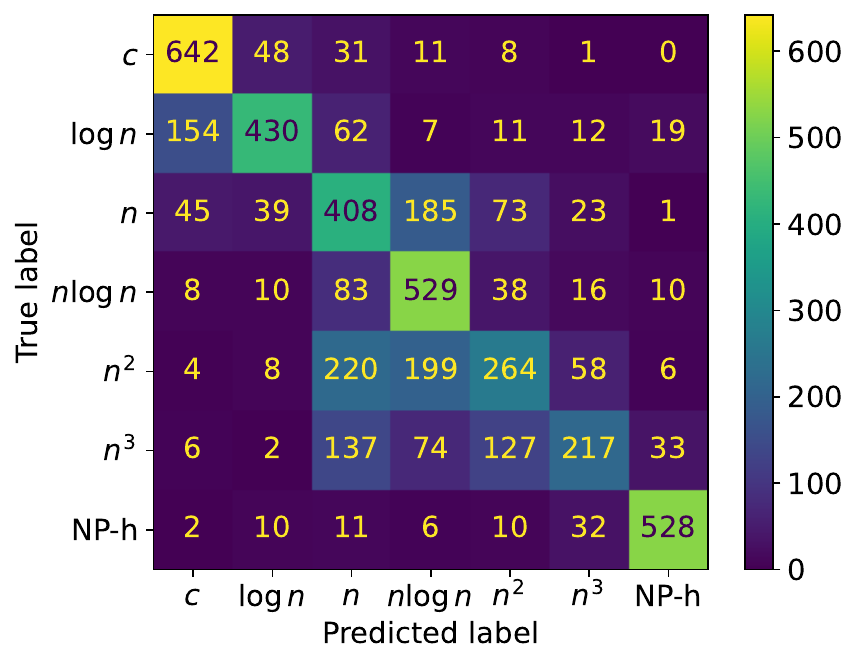}
    \includegraphics[width=0.47\textwidth]{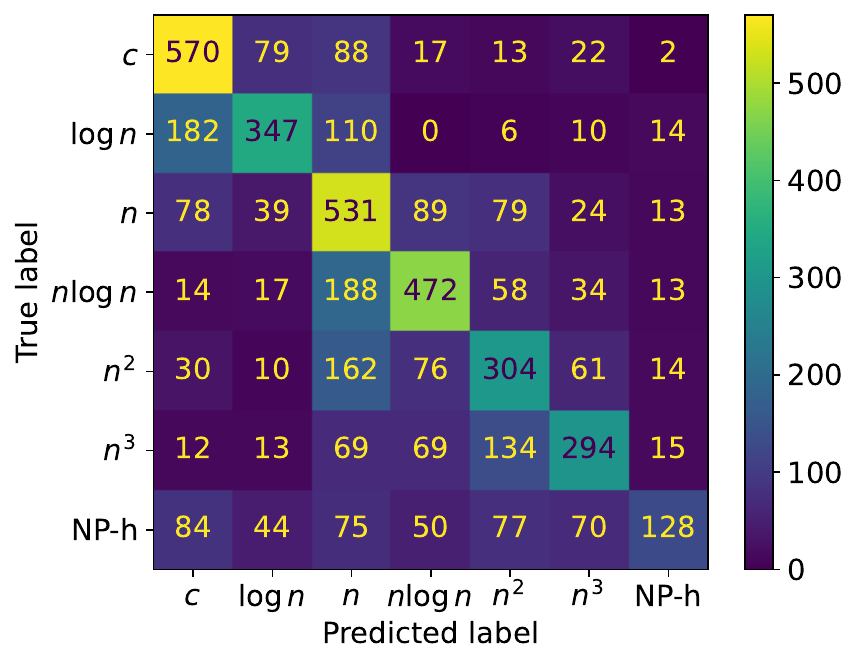}
    \caption{Confusion matrices of prediction results by PLBART on Java (left) and Python (right) datasets.}
    \label{fig:conf_PLBART}
\end{figure*}

\begin{figure*}[!htb]
    \centering
    \includegraphics[width=0.47\textwidth]{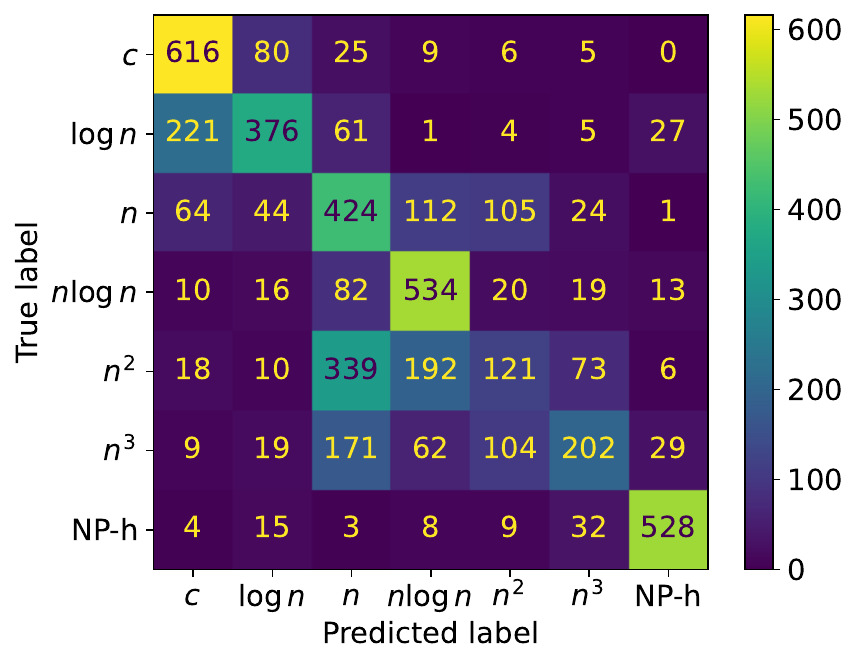}
    \includegraphics[width=0.47\textwidth]{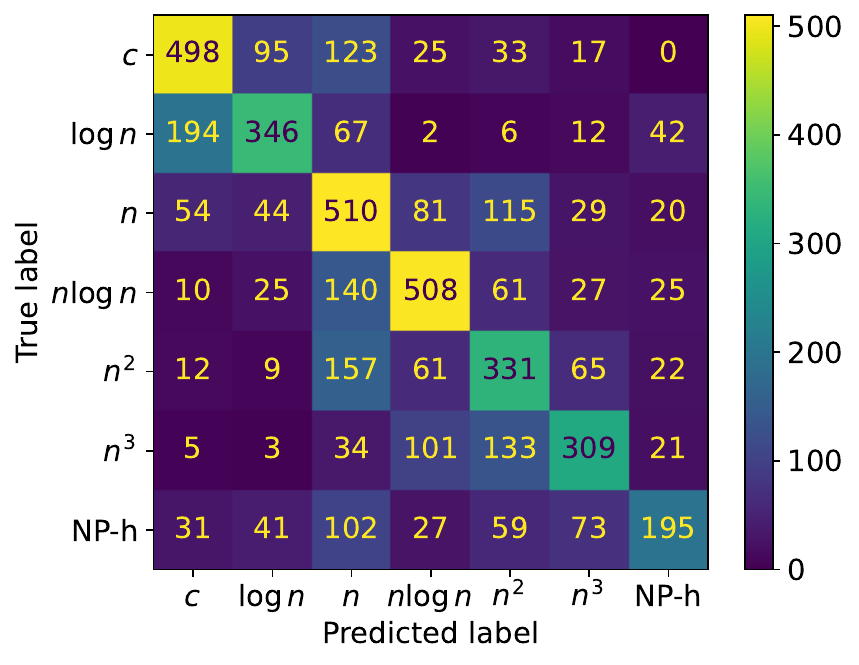}
    \caption{Confusion matrices of prediction results by UniXcoder on Java (left) and Python (right) datasets.}
    \label{fig:conf_UniXcoder}
\end{figure*}

\begin{figure*}[!htb]
    \centering
    \includegraphics[width=0.47\textwidth]{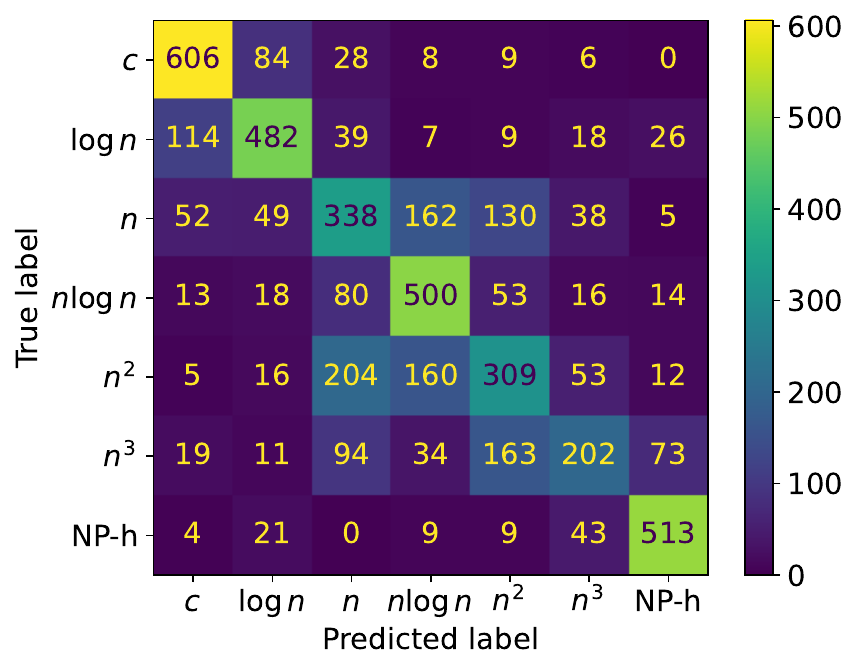}
    \includegraphics[width=0.47\textwidth]{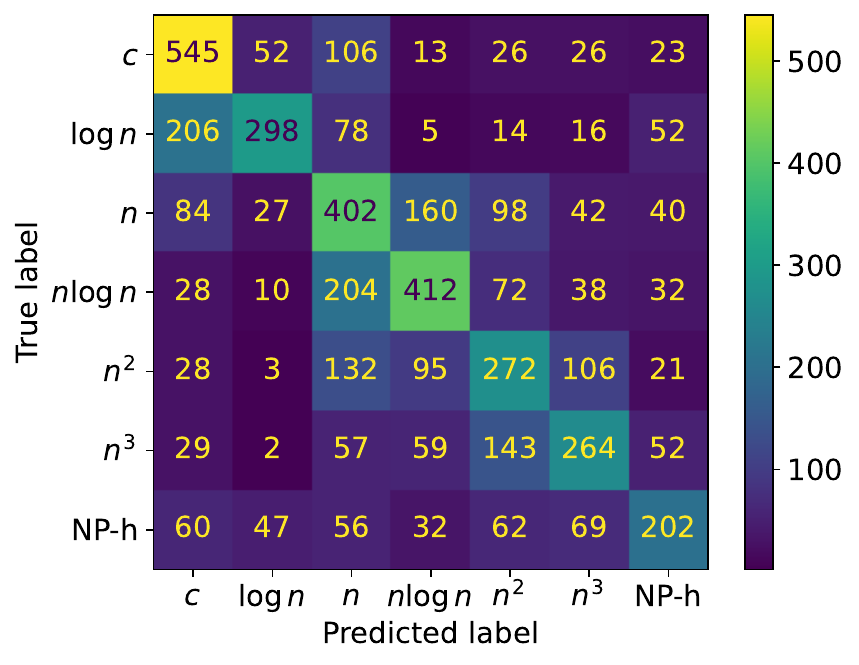}
    \caption{Confusion matrices of prediction results by CodeT5 on Java (left) and Python (right) datasets.}
    \label{fig:conf_CodeT5}
\end{figure*}

\begin{figure*}[!htb]
    \centering
    \includegraphics[width=0.47\textwidth]{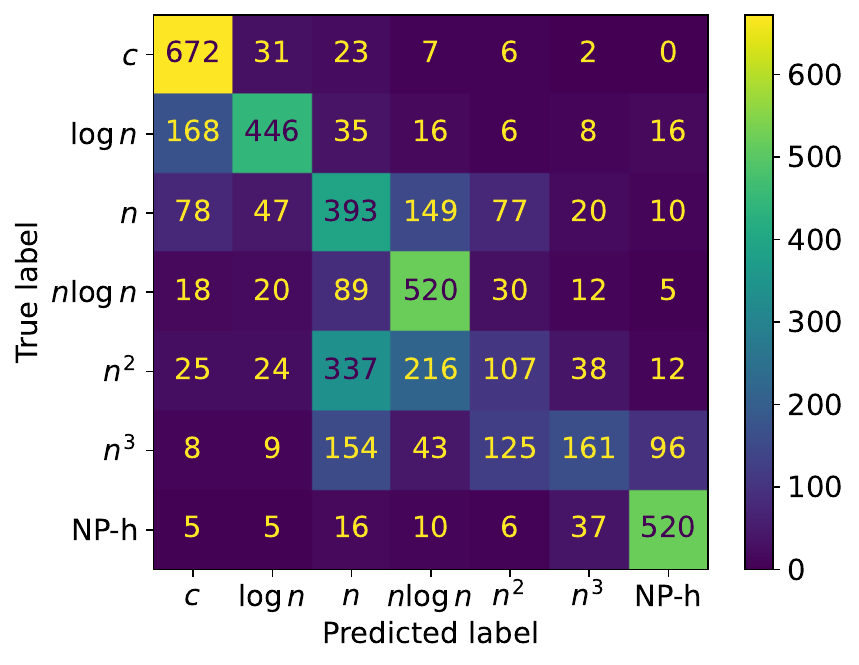}
    \includegraphics[width=0.47\textwidth]{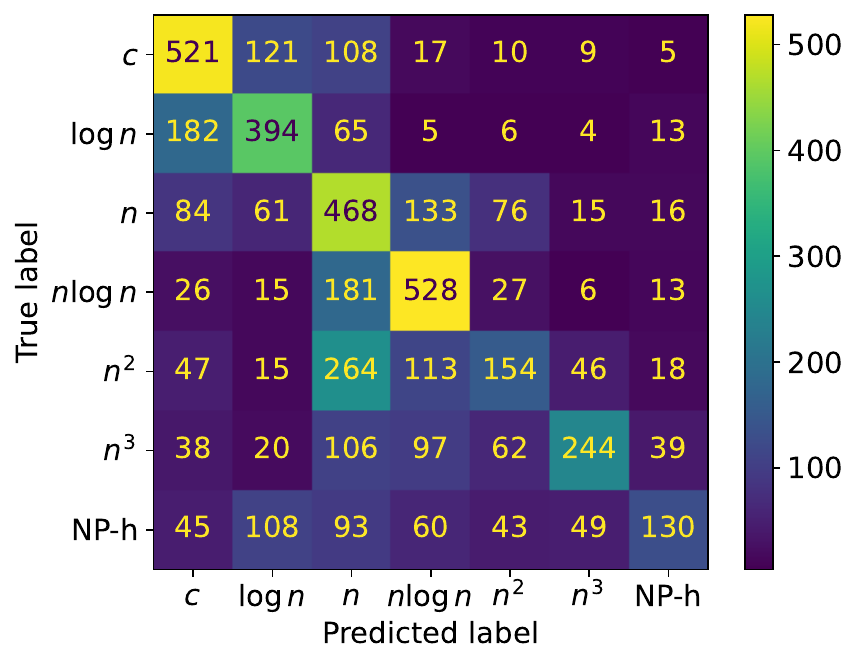}
    \caption{Confusion matrices of prediction results by CodeT5+ on Java (left) and Python (right) datasets.}
    \label{fig:conf_CodeT5P}
\end{figure*}

\end{document}